\documentclass[11pt]{article}
\usepackage[a4paper, top=16mm, text={170mm, 248mm}, includehead, includefoot, hmarginratio=1:1, heightrounded]{geometry}
\usepackage{amsmath,amssymb,mathrsfs,amsthm,tikz,shuffle,paralist}
\usepackage{dsfont}
\usepackage{color}
\usepackage{subfigure}
\definecolor{darkred}{RGB}{173,34,48}
\definecolor{lightgreen}{rgb}{0.56, 0.69, 0.19}
\definecolor{lightblue}{rgb}{0.36, 0.51, 0.71}
\definecolor{lightyellow}{rgb}{0.88, 0.61, 0.14}
\definecolor{darkgreen}{rgb}{0.6, 0.6, 0.35}
\definecolor{lightred}{rgb}{0.99, 0.36, 0.02}
\definecolor{box1}{rgb}{0.46, 0.6, 0.45}
\definecolor{box2}{rgb}{0.62, 0.56, 0.43}
\definecolor{box3}{rgb}{0.72, 0.65, 0.17}

\usetikzlibrary{snakes}
\usetikzlibrary{calc}
\usetikzlibrary{decorations}
 \usepackage{multirow}
 \usepackage{lscape}

\usepackage[all]{xypic}
\usepackage{jheppub}
\usepackage{hyperref}
\usepackage{tcolorbox}

\title{Landau-based Schubert analysis
}
\author[a,b,c]{Song He,}\emailAdd{songhe@itp.ac.cn}
\author[a]{Xuhang Jiang,}\emailAdd{xhjiang@itp.ac.cn}
\author[a,e]{Jiahao Liu,}\emailAdd{liujiahao@itp.ac.cn}
\author[d]{Qinglin Yang}\emailAdd{qlyang@mpp.mpg.de}
\affiliation[a]{CAS Key Laboratory of Theoretical Physics, Institute of Theoretical Physics, Chinese Academy of Sciences, Beijing 100190, China}
\affiliation[b]{School of Fundamental Physics and Mathematical Sciences, Hangzhou Institute for Advanced Study, UCAS, Hangzhou, 310024, China}
\affiliation[c]{Peng Huanwu Center for Fundamental Theory, Hefei, Anhui 230026, P. R. China}
\affiliation[d]{Max–Planck–Institut f\"ur Physik, Werner–Heisenberg–Institut, D–85748 Garching bei M\"unchen, Germany}
\affiliation[e]{School of Physical Sciences, University of Chinese Academy of Sciences, No.19A Yuquan Road, Beijing 100049, China}
\abstract{
We revisit the conjectural method called Schubert analysis for generating the alphabet of symbol letters for Feynman integrals, which was based on geometries of intersecting lines associated with corresponding cut diagrams. We explain the effectiveness of this somewhat mysterious method by relating such geometries to the corresponding Landau singularities, which also amounts to ``uplifting" Landau singularities of a Feynman integral to its symbol letters. We illustrate this {\it Landau-based Schubert analysis} using various multi-loop Feynman integrals in four dimensions and present an automated {\ttfamily Mathematica} notebook for it. We then apply the method to a simplified problem of studying alphabets of physical quantities such as scattering amplitudes and form factors in planar ${\cal N}=4$ super-Yang-Mills. By focusing on a small set of Landau diagrams (as opposed to all relevant Feynman integrals), we show how this method nicely produces the two-loop alphabet of $n$-point MHV amplitudes and that of the $n=4$ MHV form factors. A byproduct of our analysis is an explicit representation of any symbol alphabet obtained this way as the union of various type-$A$ cluster algebras
. }

\begin{document}
\maketitle
\section{Introduction}
Recent years have witnessed enormous progress on unravelling mathematical structures of scattering amplitudes and related quantities, especially in the remarkable theory of planar ${\cal N}{=}4$ supersymmetric Yang-Mills theory (SYM)({\it c.f.}~\cite{Arkani-Hamed:2016byb, Arkani-Hamed:2013jha}). Among others, cluster algebras~\cite{fomin2002cluster,fomin2003cluster,fomin2007cluster} have played an important role not only for its all-loop integrand~\cite{Arkani-Hamed:2016byb} but also for the functions after integration. The six- and seven-gluon amplitudes have been bootstrapped to very high loop orders~\cite{Dixon:2011pw,Dixon:2014xca,Dixon:2014iba,Drummond:2014ffa,Dixon:2015iva,Caron-Huot:2016owq,Dixon:2016nkn,Drummond:2018caf,Caron-Huot:2019vjl,Caron-Huot:2019bsq,Dixon:2020cnr,Caron-Huot:2020bkp}, where the starting point is the conjecture that the {\it symbol alphabet}~\cite{Goncharov:2010jf, Duhr:2011zq} is given by finite cluster algebras for Grassmannian $G(4,n)$ with $n=6,7$ respectively~\cite{Golden:2013xva,Golden:2014xqa}, and similar progress has been made for three- and four-point form factors~\cite{Dixon:2020bbt,Dixon:2021tdw,Dixon:2022rse,Dixon:2022xqh}. These and other considerations such as (extended) Steinmann relations~\cite{Steinmann1960a,Steinmann1960b,Caron-Huot:2016owq,Caron-Huot:2019bsq} and the so-called cluster adjacency conditions~\cite{Drummond:2017ssj, Drummond:2018caf,Drummond:2018dfd,Golden:2019kks}, have greatly restricted the space of possible multiple polylogarithmic functions (MPL). Starting from $n=8$, the cluster algebra becomes infinite and generally the symbol alphabet involves algebraic letters that go beyond cluster variables. Recent computations of two- and three-loop amplitudes~\cite{Zhang:2019vnm,He:2020vob,Li:2021bwg,Golden:2021ggj} based on~\cite{CaronHuot:2011kk, CaronHuot:2010ek, Caron-Huot:2013vda} have provided new data for $n\geq 8$, which has led to novel mathematical structures directly related to cluster algebras and positivity~\cite{Arkani-Hamed:2019rds, Drummond:2019qjk,Henke:2019hve,Drummond:2019cxm,Henke:2021avn,Ren:2021ztg,Mago:2020kmp,He:2020uhb,Mago:2020nuv,Mago:2021luw}.

More generally speaking, ${\cal N}=4$ SYM has also become an extremely fruitful laboratory for developing new tools for computing Feynman integrals, which is a subject of enormous interests by itself. There has been significant progress in studying finite, dual conformal invariant (DCI) Feynman integrals contributing to amplitudes~\cite{Drummond:2006rz, Drummond:2007aua,Drummond:2010cz,ArkaniHamed:2010gh,Spradlin:2011wp,DelDuca:2011wh,Bourjaily:2013mma,Henn:2018cdp,Herrmann:2019upk, Bourjaily:2018aeq,Bourjaily:2019hmc,He:2020uxy,He:2020lcu}. It is highly non-trivial that cluster algebras and extensions also control the symbol alphabets of individual DCI Feynman integrals (c.f.~\cite{Caron-Huot:2018dsv, Drummond:2017ssj,He:2021esx,He:2021non}). Even more surprisingly, cluster-algebraic structures have been identified and explored for the symbology of more general, non-DCI Feynman integrals in $4-2 \epsilon$ dimensions (with massless propagators)~\cite{Chicherin:2020umh,Aliaj:2024zgp}. It is natural to ask where do all these structures for symbol alphabets come from. 

In~\cite{Yang:2022gko,He:2022ctv,He:2022tph,He:2023umf}, a new method was proposed for studying the symbology of multi-loop Feynman integrals, which was based on so-called Schubert problems for geometric configurations such as intersections of lines~\cite{Hodges:2010kq,ArkaniHamed:2010gh} in momentum twistor space~\cite{Hodges:2009hk}. 
By computing cross-ratios associated with geometries for maximal cut solutions, or the \emph{leading singularities} (LS)~\cite{Bern:1994zx,Britto:2004nc,Cachazo:2008vp}, one can predict the symbol alphabet of numerous DCI Feynman integrals in the theory~\cite{Yang:2022gko,He:2022ctv}. It is very interesting that all these cluster-algebraic structures and beyond can be nicely accounted for by the Schubert analysis formulated in momentum twistor space: by introducing the line $I_{\infty}$ which breaks conformal invariance, the same kind of Schubert analysis~\cite{He:2022tph} produces symbol alphabets for Feynman integrals in $4-2\epsilon$ dimensions with various kinematics such as one-mass five-point case~\cite{Abreu:2020jxa} as well as more general integrals {\it e.g.} with internal masses~\cite{He:2023umf}. In fact, in addition to numerous examples in which Feynman integrals evaluate to MPL functions, there is also evidence that Schubert analysis can be extended to study elliptic MPL case, such as double-box integrals in $D=4$~\cite{Morales:2022csr} (see~\cite{He:2023qld} for closely related works).



More generally, it is well known that one can determine possible positions of singularities (the so-called \textit{Landau loci} in kinematic space) of a general Feynman integral by Landau analysis, which is based on a set of polynomial equations associated with cuts of the integral, \textit{i.e. Landau equations}~\cite{Landau:1959fi,Bjorken:1959fd,Nakanishi:1959,Eden:1966dnq}, and the necessary conditions for the equations to have solution. This serves as a very general method for studying analytic properties of Feynman integrals from the integrands. In recent years, there have been revived interests in studying the Landau loci and various methods and packages have been developed to systematically calculate them~\cite{Klausen:2021yrt,Mizera:2021icv,Fevola:2023kaw,Helmer:2024wax,Jiang:2024eaj,Caron-Huot:2024brh}. Landau analysis has also been used in $\mathcal{N}=4$ SYM to determine singularities to higher loops or even all loops~\cite{Prlina:2018ukf,Lippstreu:2022bib,Lippstreu:2023oio}. There have also been efforts to connect the Landau loci calculated to the symbols of MPLs~\cite{Hannesdottir:2021kpd,Dlapa:2023cvx,Jiang:2024eaj}. However, it is still unclear now how the symbol alphabets can be systematically reconstructed from Landau loci and how they can be efficiently extended to higher loops generally, though some progress has been made in the papers cited above.

In this paper, we connect these two approaches first for Feynman integrals in four dimensions, and for simplicity we illustrate the power of this {\it Landau-based Schubert analysis} in the context of ${\cal N}=4$ SYM. This has not only provided a justification from Landau analysis of the effectiveness of Schubert analysis but also given a natural extension/refinement of the Landau analysis to produce symbol alphabets. As we will see, for each Landau locus we will associate certain Schubert configurations. The geometric invariants built from such configurations provide the corresponding symbol letters associated with this locus, which contains more information than just the singularities. 
We expect this method to be general since not only the Landau analysis applies to any Feynman integrals but so does (natural extensions of) Schubert analysis when formulated in embedding space~\cite{He:2022ctv} or in Baikov representation~\cite{Baikov:1996iu,Jiang:2024eaj}. Here we choose to illustrate it using Feynman integrals in four dimensions, since the Schubert analysis takes the simplest form directly in momentum-twistor space, which makes our analysis technically much easier with clear geometric interpretations. We will apply the method to the simplest setting of planar ${\cal N}=4$ SYM, where it has been understood in~\cite{Dennen:2015bet,Dennen:2016mdk,Prlina:2017azl,Prlina:2017tvx} that each process, {\it e.g.} $n$-point MHV scattering amplitudes at two loops, only requires remarkably small number of Landau diagrams. It turns out that  this method can be used to determine the complete alphabets of two-loop $n$-point MHV and NMHV amplitudes, and as a first example we will also use it for two-loop four-point MHV form factors. 

Moreover, any alphabet obtained in this way  naturally takes the form of a union of type-$A$ cluster algebras since Schubert analysis produces (cross-ratios of) intersection points on lines. This applies to the full amplitude/form factor as well as individual Feynman integrals. In particular, for two-loop $n$-point MHV amplitude, we will see that our alphabet, which agrees with that in~\cite{Caron-Huot:2011zgw}, is always a union of $A_3$ cluster algebras. We also make comments on how algebraic letters appear in NMHV amplitudes following our method. Finally, we make discussion on future directions.

\subsection{A lightening review of the basics: Landau and Schubert analysis}
Let us begin with a lightening review of basic concepts and notations. 
Throughout the paper, most of the Feynman integrals (and amplitudes, form factors) involved will evaluate to multi-polylogarithmic (MPL) functions (with some exceptional cases which give elliptic MPL, as we will study in sec.~\ref{ellipticsec}). To encode the singularity structures, it is natural to define the symbol of MPL functions. Recall that the total differential of a weight $w$ MPL function yields a general form as
\[{\rm d}\mathcal{F}^{w}=\sum_i\mathcal{F}_i^{(w{-}1)}{\rm d}\log x_i \]
whose {\it symbol} \cite{Goncharov:2010jf, Duhr:2011zq} is iteratively defined as 
\[\mathcal{S}(\mathcal{F}^{w})=\sum_i\mathcal{S}(\mathcal{F}_i^{(w{-}1)})\otimes x_i . \]
All entries in the tensor products are called {\it symbol letters}. They are functions of kinematic variables which are closely related to the physical singularities of the quantities, and are the main interests of symbology studies. For instance, the famous four-mass box integral and its symbol reads (the dual coordinates $x_i$ are defined through $p_i:=x_{i{+}1}{-}x_i$, $x_{n{+}1}:=x_1$, and we denote $x_{i,j}^2:=(x_i{-}x_j)^2=(p_{i}{+}\cdots{+}p_{j{-}1})^2$ in this paper)
\begin{align}\label{integrand1}
    \mathcal{S}\left(\begin{tikzpicture}[baseline={([yshift=-.5ex]current bounding box.center)},scale=0.15]
                \draw[black,thick] (0,5)--(-5,5)--(-5,0)--(0,0)--cycle;
                \draw[black,thick] (1.93,5.52)--(0,5)--(0.52,6.93);
                \draw[black,thick] (1.93,-0.52)--(0,0)--(0.52,-1.93);
                \draw[black,thick] (-6.93,5.52)--(-5,5)--(-5.52,6.93);
                \draw[black,thick] (-6.93,-0.52)--(-5,0)--(-5.52,-1.93);
                \filldraw[black] (1.93,6) node[anchor=west] {{$j{-}1$}};
                \filldraw[black] (0.52,6.93) node[anchor=south] {{$i$}};
                \filldraw[black] (1.93,-1) node[anchor=west] {{$j$}};
                \filldraw[black] (0.52,-1.93) node[anchor=north] {{$k{-}1$}};
                \filldraw[black] (-6.93,6) node[anchor=east] {{$l$}};
                \filldraw[black] (-5.52,6.93) node[anchor=south] {{$i{-}1$}};
                \filldraw[black] (-6.93,-1) node[anchor=east] {{$l{-}1$}};
                \filldraw[black] (-5.52,-1.93) node[anchor=north] {{$k$}};
            \end{tikzpicture}\right)= \frac1{2\Delta
            } \biggl(v\otimes \frac{z}{\bar{z}} 
            +u\otimes \frac{1-\bar{z}}{1-z}
            \biggr)
\end{align}
with the definition of cross-ratios $z,\bar{z}$ and square root $\Delta$:
\begin{eqnarray}
\label{delta}
&u=\frac{x_{i,j}^2 x_{k,l}^2}{x_{i,k}^2 x_{j,l}^2}=z \bar{z},\ v=\frac{x_{i,l}^2 x_{j,k}^2}{x_{i,k}^2 x_{j,l}^2}=(1-z)(1-\bar{z}),\nonumber\\
&\Delta:=\sqrt{(1-u-v)^2-4 u v}.
\end{eqnarray}
Here $\Delta$ is also referred to as the leading singularity \cite{Arkani-Hamed:2010pyv} of the four-mass box integral and the symbol letters involve $z, \bar{z}, 1-z, 1-\bar{z}$ (for simplicity we have suppressed the dependence on four dual points $x_i, x_j, x_k, x_l$). 
As we have reviewed, various methods have been developed in the past decade both for $\mathcal{N}=4$ SYM theory \cite{Spradlin:2011wp,Dixon:2011pw,Drummond:2014ffa,Golden:2014pua,Dixon:2016nkn,Drummond:2017ssj,Prlina:2018ukf,Drummond:2019cxm,Caron-Huot:2020bkp,He:2020vob,Mago:2020kmp,Mago:2020nuv,Mago:2021luw,Dixon:2021tdw,He:2020uxy,He:2020lcu,He:2021esx,He:2021non,He:2021eec,He:2022tph,Wilhelm:2022wow,Morales:2022csr,He:2023qld} and more generally \cite{Abreu:2017mtm,Broedel:2018iwv,Chicherin:2020umh,Abreu:2021vhb,Gong:2022erh,He:2023umf,Dlapa:2023cvx,Chen:2023kgw,Jiang:2024eaj}. Next we review the Landau analysis and the Schubert analysis in momentum twistor space. 

\paragraph{Landau equations and singularities}
Singularity in kinematics space for certain Feynman integrals was understood formally through Landau analysis and related Landau equations \cite{Landau:1959fi,Eden:1966dnq}. As a brief review, suppose we are dealing with a standard Feynman integral after Feynman parametrization
\begin{equation}
    \int\prod_{i=1}^L{\rm d}^D\ell_i\int_{\mathcal{C}}{{\rm d}^\nu\alpha}\frac{\mathcal{N}}{\mathcal{D}^\nu},
\end{equation}
where the denominator reads $\mathcal{D}=\sum_{i=1}^\nu\alpha_i(q_i^2-m_i^2)$. $q_i$ is the loop momentum running through the $i$th propagator. $m_i^2$ is the mass of the propagator which always reads $0$ in this work. $\mathcal{N}$ is the numerator involving $\ell_i$ and external $p_i$. And the integration is performed through the contour $\mathcal{C}=\{\sum_i\alpha_i=1,\ \alpha_i\geq0\}$. Singularities of this integral can only appear when surface $\mathcal{D}=0$ pinches the integration contour, and can be determined by Landau equations as the following \cite{Eden:1966dnq}
\begin{align}
    &\alpha_i(q_i^2-m_i^2)=0\ \ \ \ \ \ \ \ \text{(cut condition)}\label{cut},\\
    &\sum_{i\in \text{each loop}}\alpha_iq_i^\mu=0\ \ \ \text{(pinch condition)}\label{pinch}, 
\end{align}
{\it i.e.} physical singularities are conditions constraining external kinematics $p_i$, such that  equations \eqref{cut} and \eqref{pinch} have non-trivial solutions for $\alpha_i$. They are also called \textit{Landau loci} of the integral. Moreover, a branch of solution is called {\it leading Landau singularity} or leading singularity if we have all $\alpha_i\neq0$, and we will call a singularity {\it sub-leading} if the associated solution for $\alpha_i$ consists of at least one $\alpha_i=0$, {\it etc.}. In another word, sub$^k$-leading Landau singularities of the integral are leading singularities of all sub-topologies of the original integral, obtained by shrinking $k$ of its propagators. When performing Landau analysis for an individual integral, we should not only check its leading singularities, but also all sub$^k$-leading singularities, in order to guarantee all possible physical singularities are covered.

We now can see that all singularities and square roots of four-mass box emerge from Landau loci. Landau equations of the integral read
\begin{align}
    &\alpha_i (x_{AB}-x_i)^2=\alpha_j (x_{AB}-x_j)^2=\alpha_k (x_{AB}-x_k)^2=\alpha_l (x_{AB}-x_l)^2=0\label{eq:boxcut},\\
    &\alpha_i(x_{AB}-x_i)^\mu+\alpha_j(x_{AB}-x_j)^\mu+\alpha_k(x_{AB}-x_k)^\mu+\alpha_l(x_{AB}-x_l)^\mu=0.
\end{align}
On the support of $\alpha_i\neq0$, {\it i.e.} the leading singularity solution branch, $x_{AB}$ is fully localized by the four cut conditions (which is just the Schubert solution in the next paragraph), and the pinch condition can be translated to 
\begin{equation}
    \left(\begin{matrix}
        0&x_{ij}^{2} & x_{ik}^2 & x_{il}^{2} \\
       x_{ij}^{2} & 0 & x_{jk}^2 & x_{jl}^2\\
       x_{ik}^2 & x_{jk}^2&0& x_{kl}^2\\
       x_{il}^2& x_{jl}^2&x_{kl}^2&0
    \end{matrix}\right)\cdot \left(\begin{matrix}
        \alpha_1\\\alpha_2\\\alpha_3\\\alpha_4
    \end{matrix}\right):=Q\cdot{\bf \alpha}=0,
\end{equation}
and the equations have non-trivial solution if and only if $\det Q\propto \Delta^2=0$, which is just the leading singularity for four-mass box as we have seen. We note here that when $\det Q=0$, there indeed exists a solution which satisfies that \textit{all} $\alpha_{i}\neq 0$. This is the condition for leading singularities. Similar analysis can be applied to sub-leading singularities by setting any of the $\alpha_i=0$. Those solution branches are equivalent to leading Landau loci of sub-topologies for the box (triangles, bubbles, {\it etc.}), and all Landau singularities computed are just factors of $u$ and $v$ in \eqref{delta}.

We should also emphasize that $\Delta=0$ has a geometrical meaning as pinching of two solutions from the cut condition \eqref{eq:boxcut}. As can be proved from embedding formalism \cite{He:2023umf}, as a quadratic equation, discriminant of \eqref{eq:boxcut} after proper parametrization reads exactly $\Delta^2$. In another word, we always have relation $(x_{AB}^{(1)}-x_{AB}^{(2)})^2\propto \Delta^2$ for box configuration, or any configuration equivalent to one-loop box. This will be our starting point for geometrizing Landau loci and uplifting loci to letters in the next section.

Some comments are in order. Firstly, Landau analysis can only work out physical singularities for the integral, but not their letters. One can see that in Eq.~\eqref{integrand1}, besides $u$ and $v$ which are generated from Landau equations, we also have two other letters $\{\frac{z_{i,j,k,l}}{\bar{z}_{i,j,k,l}},\frac{1{-}z_{i,j,k,l}}{1{-}\bar{z}_{i,j,k,l}}\}$, which yield physical singularities $u$ and $v$ as well. Generally speaking, for physical singularities $W_i$ and their multiplicative combinations $\prod_i W_i$, we always have more than one way to rewrite it as 
\[\frac{a{+}\sqrt{\Delta}}{a{-}\sqrt{\Delta}},\ \ (a{+}\sqrt{\Delta})(a{-}\sqrt{\Delta})\propto \prod_i W_i\]
with certain square root $\sqrt{\Delta}$ from Landau analysis, resulting in different letters. In this work, with the help of Schubert analysis, we will generalize Landau analysis and get symbol letters exactly. Secondly, although Landau equations are fully determined by the propagator structure of the integral, existence of numerator $\mathcal{N}$ will also affect the solution of the equation, since certain solution for $\ell_i$ and $p_i$ may result in $\mathcal{N}=0$, and the singularity is in fact absent. We will see this point when we consider chiral numerators in MHV, NMHV amplitudes. Finally, we omit the so-called second type Landau singularities \cite{FLNP:1962,Eden:1966dnq}, which is related to the ultraviolet behavior of the integral when $\ell\to\infty$. However, since we will restrict ourselves in $\mathcal{N}=4$ SYM theory and dual conformal invariant world, this kind of singularities is totally absent.

\paragraph{Schubert analysis in momentum twistor space}
Throughout this note, we will adopt momentum twistor variables \cite{Hodges:2009hk,Mason:2009qx} to represent kinematics, whose definition is
\[
 Z_{i}^{a}:=(\lambda_{i}^{\alpha},x_{i}^{\alpha\dot{\alpha}}\lambda_{i\alpha}),
 \]
where we use dual coordinates to represent the external momenta  $p_i^{\alpha\dot{\alpha}}:=(x_{i{+}1}{-}x_i)^{\alpha\dot{\alpha}}=\lambda_i^\alpha\tilde{\lambda}_i^{\dot{\alpha}}$. Under the momentum-twistor representations, squared distance of any two dual coordinates reads $(x_i-x_j)^2=(p_i{+}\cdots{+}p_{j{-}1})^2=\frac{\langle i{-}1ij{-}1j\rangle}{\langle i{-}1i\rangle\langle j{-}1j\rangle}$, with Pl\"ucker variables $\langle ijkl\rangle:=\epsilon_{ABCD}Z_i^AZ_j^BZ_k^CZ_l^D$ and 2-brackets $\langle ij\rangle:=\epsilon_{AB}\lambda_i^{A}\lambda_j^B$. This provides a correspodence between each dual coordinates $x_i$ and each line (bitwistor) $(i{-}1i):=\epsilon_{ABCD}Z_{i{-}1}^AZ_i^B$ in momentum twistor space. Therefore, by associating loop momenta $\ell$ with line $(AB)$ in momentum twistor space, propagator $(\ell-x_i)^2$ for any integral can also be transformed to $\frac{\langle ABi{-}1i\rangle}{\langle AB\rangle\langle i{-}1i\rangle}$. Dual conformal symmetry \cite{Drummond:2006rz,Bern:2006ew,Drummond:2007aua} is converted to $SL(4)$ invariance in momentum twistor space, indicating that all dependence of 2-brackets will be canceled in the final result of integrands or integrated results. We will also use shorthand $(ijk):=\epsilon_{ABCD}Z_i^AZ_j^BZ_k^C$ and $\bar{i}:=(i{-}1ii{+}1)$ to represent planes later on for simplicity. The intersection twistor of a line and a plane, which is thus a point, will be $(ij)\cap(lmn)$ and $(ijk)\cap(lmn)$ is the intersection bitwistor (a line) of two planes, {\it etc.} The line lies at infinity is defined as
\begin{equation}
    I_{\infty}=\left(\begin{array}{cc}
        0 & 0 \\
        0 & 0 \\
        0 & 1 \\
        1 & 0 
    \end{array}\right).
\end{equation}
The last important shorthand is $\langle a(bc)(de)(fg)\rangle:=\langle (abc)\cap(ade) fg\rangle$. The physical quantities we will study in the follows are all functions of these momentum-twistor variables.
 
\begin{figure}[t]
    \centering
    \includegraphics[width=0.35\linewidth]{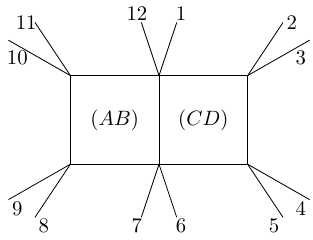}
    \caption{The fully massive double-box.}
    \label{fig:12pdb}
\end{figure}

\begin{figure}[t]
    \centering
    \includegraphics[width=0.65\linewidth]{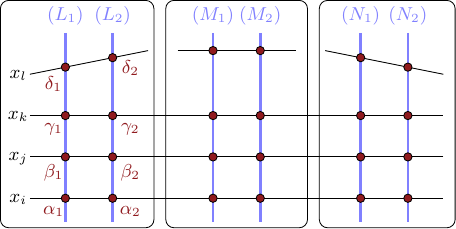}
    \caption{Combining Schubert configurations for triple of four-mass boxes when analyzing MPL letters for double-box integral}
    \label{fig:schuberteg}
\end{figure}

Finally, we review Schubert analysis in momentum twistor space \cite{Yang:2022gko}.
The basic idea of Schubert analysis is to geometrize on-shell solutions for loop momenta of target integral by lines in momentum twistor space, both for integral itself and its sub-topologies\footnote{Note that in previous study \cite{Yang:2022gko}, sub-topologies of certain $L$-loop integral in Schubert story contain not only $L$-loop integrals but also lower-loop integrals. In fact, all these sub-topologies are actually determined by Landau equation system, {\it i.e.} they correspond to sub-leading Landau equation systems from the top integral, as we will see.}. On-shell conditions $q_i^2{-}m_i^2=0$ from propagators result in intersection configurations formed by loop momenta solutions $(AB)_i$ and external kinematics lines $(i{-}1i)$, and all symbol letters are generated from cross-ratios of the intersection points. As an illustration, we present the $A_3$ configuration in \cite{Golden:2013xva} when we consider the double-box integral (Fig.~\ref{fig:12pdb}) and its MPL symbol letters (Fig.~\ref{fig:schuberteg}). In Fig.~\ref{fig:schuberteg}, each boxed Schubert configuration is from a four-mass box, which is a sub-topology of double-box integral. $L_1$, $L_2 $ (and $M_i$, $N_i$) are two Schubert solutions from four-mass box $I_4(i,j,k,l)$, and four intersections on each solution gives symbol letters
\begin{align}
\frac{[\alpha_1,\beta_1][\gamma_1,\delta_1]}{[\alpha_1,\gamma_1][\beta_1,\delta_1]}=z_{i,j,k,l},\frac{[\alpha_1,\delta_1][\gamma_1,\beta_1]}{[\alpha_1,\gamma_1][\beta_1,\delta_1]}=1{-}z_{i,j,k,l}\nonumber\\
\frac{[\alpha_2,\beta_2][\gamma_2,\delta_2]}{[\alpha_2,\gamma_2][\beta_2,\delta_2]}=\bar z_{i,j,k,l},\frac{[\alpha_2,\delta_2][\gamma_2,\beta_2]}{[\alpha_2,\gamma_2][\beta_2,\delta_2]}=1{-}\bar{z}_{i,j,k,l}
\end{align}
with $[Z_i,Z_j]:=\langle ij I\rangle$ for some reference bitwistors. And one can see that these four letters exactly match the result in \eqref{integrand1}. 


Furthermore, besides cross-ratios from the blue lines,  on $x_i$ $x_j$ and $x_k$ we can construct $9$ independent cross-ratios, consisting of three rational letters proportional to $\mathcal{G}^{a,b}_{a,b}$ for $\{a,b\}\in\{\{i,j\},\{j,k\},\{i,k\}\}$, as well as six odd letters as
\begin{align}\label{odds12}
    &\left\{\frac{\mathcal{G}^{a}_{b}+\sqrt{-\mathcal{G}_{a,b}^{a,b}\mathcal{G}}}{\mathcal{G}^{a}_{b}-\sqrt{-\mathcal{G}_{a,b}^{a,b}\mathcal{G}}}\right\},\, \{a,b\}\in\{\{i,j\},\{j,k\},\{i,k\}\},\\
&\left\{\frac{\mathcal{G}^{a,b}_{a,c}+\sqrt{\mathcal{G}_{a,b}^{a,b}\mathcal{G}_{a,c}^{a,c}}}{\mathcal{G}^{a,b}_{a,c}-\sqrt{\mathcal{G}_{a,b}^{a,b}\mathcal{G}_{a,c}^{a,c}}}\right\},\,  \{a,b,c\}\in\{\{i,j,k\},\{j,k,i\},\{k,i,j\}\},
\end{align}
where $\mathcal{G}$ reads the Gram determinant of fully massive hexagon, and $\mathcal{G}_B^A:=\det\{x_{ij}^2\}$ with $i\in \{2,4,6,8,10,12\}{-}A$, $j\in \{2,4,6,8,10,12\}{-}B$. They are all reasonable candidates for the third entries of $12$-point double-box integral \cite{Morales:2022csr}, and the configurations on $x_i$, $x_j$ or $x_k$ enjoy natural $A_3$ cluster structures \cite{Golden:2013xva}.  Comparing to Landau analysis, Schubert analysis offers us further information, {\it i.e.} we get exactly the letters instead of only singularities. Although this method works in many non-trivial examples, the mechanism of Schubert analysis lacks explanation. In this work we will try to relate the method with Landau analysis, and see that Schubert analysis work out symbol letters for amplitudes/integrals reasonably following Landau equations. 

\section{Schubert analysis based on Landau analysis}\label{sec:2}

In this section we revisit the method of Schubert analysis for generating alphabets of Feynman integrals from the point of view of Landau analysis. Usually, the terminology of ``Landau analysis" concerns conditions\footnote{These conditions are usually sufficient and we do not know whether they are necessary or not in general,  that is, nontrivial singularities may be missed.}, {\it i.e.} Landau loci, for Landau equations to have non-trivial solutions as mentioned in the last section. However, from our point of view, it is also important to find the solution space of Landau equations when nontrivial solutions exist, which turns out to be the starting point of Schubert analysis. In this section, we will describe this method in detail and show that in addition to symbol letters which are given by Landau loci, the so-called algebraic letters and the ``mixing" in Schubert analysis naturally arise from the way we pinch these solutions in one-dimensional solution space. Furthermore, this point of view can be generalized even to cases involving elliptic integrals.

\subsection{Pinching solutions of Landau equations and Schubert analysis}
\begin{figure}[htbp]
    \centering
    \includegraphics[width=0.35\linewidth]{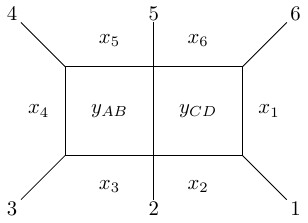}
    \caption{6-point double-box integral.}
    \label{fig:doublebox}
\end{figure}

To state the idea clearly, we stick to a simple example and state the general rules therein. Let us consider Landau problems in the following two-loop six-point double-box integral which is depicted in Fig.~\ref{fig:doublebox}.
The definition of dual points follows from the introduction. Then we can list the Landau equations of this problem.
\begin{equation}\label{eq:landufor6pdb}
\begin{aligned}
    \mathcal{D}=&\alpha_{1}(y_{CD}-x_1)^2+\alpha_{2}(y_{CD}-x_2)^2+\alpha_{6}(y_{CD}-x_6)^2 \\
    &\hspace{4em}+\alpha_{3}(y_{AB}-x_3)^2+\alpha_{4}(y_{AB}-x_4)^2+\alpha_{5}(y_{AB}-x_5)^2 +\beta(y_{AB}-y_{CD})^2,\\
    \text{(cuts): }&\alpha_i\frac{\partial\mathcal{D}}{\partial\alpha_{i}}=0;\quad\quad \beta\frac{\partial\mathcal{D}}{\partial\beta}=0. \\
    \text{(pinch): }&\frac{1}{2}\frac{\partial\mathcal{D}}{\partial y_{AB}}=\alpha_{3}(y_{AB}-x_3)+\alpha_{4}(y_{AB}-x_4)+\alpha_{5}(y_{AB}-x_5)+\beta(y_{AB}-y_{CD})=0, \\
    \text{(pinch): }&\frac{1}{2}\frac{\partial\mathcal{D}}{\partial y_{CD}}=\alpha_{1}(y_{CD}-x_1)+\alpha_{2}(y_{CD}-x_2)+\alpha_{6}(y_{CD}-x_6)+\beta(y_{CD}-y_{AB})=0.
\end{aligned}
\end{equation}
We note that, when some $\alpha_{i}=0$ or $\beta=0$, it corresponds to the subsectors\footnote{Subsectors are defined by pinching some propagators of a given Feynman integral, they are also called subtopologies.} by pinching corresponding propagators. So equivalently all subsectors of the above diagram need to be considered to get the full solution space for $y_{AB}$ or $y_{CD}$.
We will focus on $y_{AB}$, the left loop momentum of the double-box, without loss of generality since the right loop $y_{CD}$ is totally symmetric under the exchange of external legs.

\begin{figure}[htbp]
    \centering
    \includegraphics[width=0.25\linewidth]{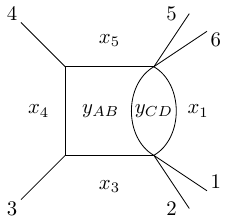}
    \caption{One box-bubble subtopology for double-box diagram}
    \label{fig:6pdbsub_bb}
\end{figure}
\begin{figure}[htbp]
    \centering
    \includegraphics[width=0.35\linewidth]{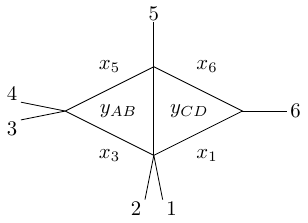}
    \caption{The double-triangle subtopology of the double-box}
    \label{fig:6pdbsub_dt}
\end{figure}

We start with the subsectors which are simpler and then study the top sector. Most of its subsectors are trivial since after enough pinches the Landau equations of these diagrams will be equivalent to one-loop bubbles, triangles or boxes\footnote{One-loop tadpoles will directly be 0 due to the scaleless property.}. For instance, sub$^2$-leading diagram Fig.~\ref{fig:6pdbsub_bb} with two propagators shrunk results in Landau equation
\begin{equation}\label{eq:box1}
    {\color{lightyellow} \mathrm{L}_{3451}}: \,\langle AB23\rangle=\langle AB34\rangle=\langle AB45\rangle=\langle AB61\rangle=0,
\end{equation}
where $\langle AB61\rangle$ comes from the pinch condition for $y_{CD}$, and other three are from cut conditions. 
A similar logic applies to sub$^2$-leading sector Fig.~\ref{fig:6pdbsub_dt}, and its Landau equation reads
\begin{equation}\label{eq:box3}
    {\color{lightblue} \mathrm{L}_{3561}}: \, \langle AB23\rangle=\langle AB45\rangle=\langle AB56\rangle=\langle AB61\rangle=0.
\end{equation}
where $\langle AB61\rangle=\langle AB56\rangle=0$ are from pinch condition of $y_{CD}$ as well. For a general discussion of pinch conditions in loop-by-loop case, see App.~\ref{app:triangle}. Following this analysis, for any two-loop integral, all possible one-loop bubbles, triangles or boxes Landau equations formed by any 2,3 or 4 of its dual points always appear as sub$^k$-leading Landau equation for the top integral, and we should always include their leading Landau loci $\{x_{ij}^2\ \& \ \mathbf{Gram}_{a,b,c,d}\}$ as part of our singularities
, where we use the notation $\mathbf{Gram}_{A}:=\det\{\langle i{-}1ij{-}1j\rangle\}$ for all $i,j\in A$.

\begin{figure}
    \centering
    \includegraphics[width=0.35\linewidth]{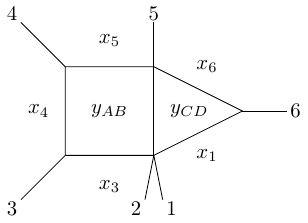}
    \caption{One box-triangle subtopology for double-box in Fig.~\ref{fig:doublebox}.}
    \label{fig:6pdbsub_bt}
\end{figure}

\begin{figure}[htbp]
    \centering
    \includegraphics[width=0.35\linewidth]{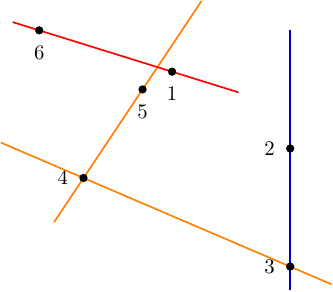}
    \caption{The Schubert configuration for one-massive-leg box: This is a configuration for $\langle AB23\rangle=\langle AB34\rangle=\langle AB45\rangle=\langle AB61\rangle=0$. The lines are colored to indicate that they are not in the same plane. We can find two solutions for $(AB)$ to intersect all four lines.}
    \label{fig:schubert_oneloop}
\end{figure}

One of the non-trivial sub-sectors will be box-triangle diagrams by pinching one propagator. Let us take the subsector in Fig.~\ref{fig:6pdbsub_bt} as an example.
Following the discussion for general triangles in App.~\ref{app:triangle}, we can easily write down the pinch condition for $y_{CD}$,
\begin{equation}\label{eq:pinchcond}
    \langle AB56\rangle=\langle AB61\rangle=0.
\end{equation}
Combining the above constraints with the cut condition for $y_{AB}$, we will arrive at the following five conditions for $y_{AB}$:
\begin{equation}\label{eq:fiveconstraints}
   {\color{lightred} \mathrm{L}_{34561}}:\, \langle AB23\rangle=\langle AB34\rangle=\langle AB45\rangle=\langle AB56\rangle=\langle AB61\rangle=0.
\end{equation}
It is obvious that ${\color{lightred} \mathrm{L}_{34561}}$ in Eq.~\eqref{eq:fiveconstraints} cannot be satisfied for general kinematics when $y_{AB}$ is in four dimensions, unless there is extra singular condition for external kinematics, which is part of the Landau loci. It can be derived by, {\it e.g.} solving the first four conditions (which is called a Schubert problem as first noted in \cite{Arkani-Hamed:2010pyv}\footnote{This name originates from the Schubert calculus which is introduced by Schubert to solve classic counting problems of projective geometry. Hereafter we will call a set of such conditions as a Schubert configuration. The problem of solving such conditions will be called a Schubert problem and corresponding solutions are named Schubert solutions.}) and then substitute this solution into the fifth one. The solution of the first four conditions can be visualized in Fig.~\ref{fig:schubert_oneloop} in momentum twistor space and it is solved to be
\begin{equation}\label{eq:onemassbox_sol}
    (AB)=(61)\cap(234)4, \quad (AB)=(61)\cap(345)3.
\end{equation}
After substituting above two solutions into the last condition we will get two Landau loci
\begin{equation}\label{eq:5pGram}
    \langle(61)\cap(234)456\rangle=\langle 1456\rangle\langle 2346\rangle, \quad \langle(61)\cap(345)356\rangle=\langle 1356\rangle\langle 3456\rangle.
\end{equation}
We will encounter many other different configurations for different Schubert problems and this is handled by a \texttt{Mathematica} notebook presented in the ancillary files. We should mention here that this Landau locus is actually the Gram determinant formed by the five dual coordinates
\[\mathbf{Gram}_{1,3,4,5,6}=2\langle1356\rangle\langle1456\rangle\langle2346\rangle\langle3456\rangle.\]
Obviously, a similar analysis can be applied to general box-triangle integrals with more massive external corners, and we conclude that box-triangle sub-topologies always yield $\mathbf{Gram}_{a,b,c,d,e}$ Landau loci for the top integral. In next section, we will see that symbol letters $\mathbf{Gram}_{a,a{+}1,b,b{+}1,c}$ appear in 2-loop MHV amplitudes following a similar logic.

\begin{figure}[htbp]
    \centering
    \includegraphics[width=0.4\linewidth]{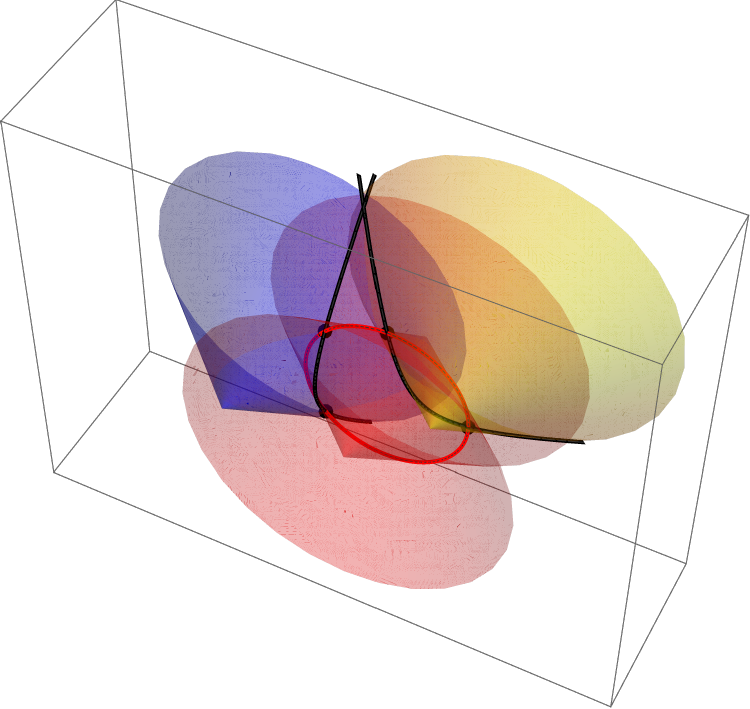}
    \caption{To display the solutions in a diagram, here we suppose the loop momentum $y_{AB}$ is a three dimension vector living in $SO(2,1)$. Then we first need two common conditions ($(y_{AB}-x_1)^2=(y_{EF}-x_2)^2=0$) which are indicated by the red cones. Their intersection gives a one-dimension subspace which is the red circle (In $SO(3,1)$ Minkowski space, we need three such conditions just as mentioned in the main context). Then we put on two different conditions $(y_{EF}-x_3)^2=0$ (the blue) and $(y_{EF}-x_4)^2=0$ (the yellow). Each interacts one of the red cones with a parabola (drawn in black line). And each parabola intersects the red circle with two black points. Landau loci are given by the pinches between solutions (the black dots) in the one-dimension subspace (the red circle).}
    \label{fig:solution}
\end{figure}

Now let us look into the procedure to solve singularities in \eqref{eq:5pGram} more carefully via a geometrical viewpoint. Actually, this procedure can be viewed as the pinches of solutions in the solution space of $y_{AB}$, one solution from \eqref{eq:box1}, and another from \eqref{eq:box3}. These two conditions only differ from each other by one condition. Therefore $\eqref{eq:5pGram}=0$ indicates that solutions for \eqref{eq:box1} can also solve \eqref{eq:box3}, {\it i.e.} solutions from the two conditions pinch together.    
Then for now, we can interpret our calculation of Landau loci as the pinches in a one-dimension solution space for $y_{AB}$ which is determined by $\langle AB23\rangle=\langle AB34\rangle=\langle AB45\rangle=0$. This is visualized in Fig.~\ref{fig:solution}.
Four solutions of $AB$ from the two problems read 
\begin{equation}\label{eq:fourso}
    \begin{aligned}
        &{\color{lightyellow} \mathrm{L}_{3451}}: \quad l_1=[(61)\cap(234)4], \, l_2=[(61)\cap(345)3]; \\
        &{\color{lightblue} \mathrm{L}_{3561}}: \quad l_3=[(23)\cap(456)6], \, l_4=[(23)\cap(561)5].
    \end{aligned}
\end{equation}
and it can be easily checked that we have
\begin{equation}\label{eq:locus}
(l_1\cdot l_3)(l_1\cdot l_4)(l_2\cdot l_3)(l_2\cdot l_4)=\frac14\langle 1236\rangle^2\mathbf{Gram}_{1,3,4,5,6}^2
\end{equation}
with the notation $(L_1\cdot L_2)=\langle1234\rangle$ for $L_1=(12)$, $L_2=(34)$. In momentum space, $(L_1\cdot L_2)$ stands for the square of distance for the two solutions $L_1$ and $L_2$. Any pairs of solutions pinching in \eqref{eq:locus} will result in singularity $\mathbf{Gram}_{1,3,4,5,6}\to0$.

\begin{figure}
    \centering
    \includegraphics[width=0.5\linewidth]{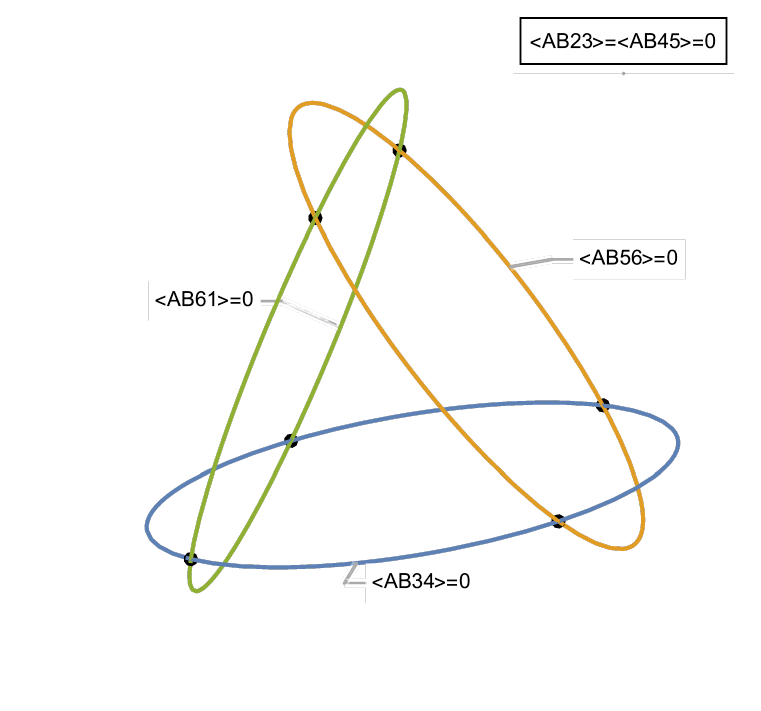}
    \caption{The solutions of $y_{AB}$ in the space under constraint $\langle AB23\rangle=\langle AB45\rangle=0$. Three pairs of solutions come from three different configurations. The top pair corresponds to ${\color{lightblue} \mathrm{L}_{3561}}$, the left bottom pair corresponds to ${\color{lightyellow} \mathrm{L}_{3451}}$ and the right bottom pair corresponds to ${\color{lightgreen} \mathrm{L}_{3456}}$. Each of the three problems corresponds to the Landau conditions of some subsectors of double-box. Note that in this figure we just display in one way how these three circles intersect with each other, there are other ways certainly.}
    \label{fig:solution2}
\end{figure}

We should comment that taking the constraint ${\color{lightred} \mathrm{L}_{34561}}$ as pinching of solutions for ${\color{lightyellow} \mathrm{L}_{3451}}$ and ${\color{lightblue} \mathrm{L}_{3561}}$ is not our only choice. Firstly, the choice of these two configurations ${\color{lightyellow} \mathrm{L}_{3451}}$ and ${\color{lightblue} \mathrm{L}_{3561}}$ is not made in a random way, since these two conditions correspond to two sub-topologies in this double-box, as we mentioned. And secondly there are many different combinations of sub-topologies in this double-box which can result in the same constraints as ${\color{lightred} \mathrm{L}_{34561}}$. For instance, we can also view ${\color{lightred} \mathrm{L}_{34561}}$ as a combination of ${\color{lightyellow} \mathrm{L}_{3451}}$ and ${\color{lightgreen} \mathrm{L}_{3456}}$  
\begin{equation}\label{eq:box2}
    {\color{lightgreen} \mathrm{L}_{3456}}: \,\langle AB23\rangle=\langle AB34\rangle=\langle AB45\rangle=\langle AB56\rangle=0.
\end{equation}
And ${\color{lightgreen} \mathrm{L}_{3456}}$ corresponds to another box-bubble which can be got by pinching $\langle CD61\rangle$ in the box-triangle of Fig.~\ref{fig:6pdbsub_bt}. Actually, we can depict the three problems and their solutions in Fig.~\ref{fig:solution2},
where any two black intersections shared by two circles are solutions from one box Schubert configuration. 

From the diagram, the Schubert intersection configuration naturally arises: they acturally correspond to ``a combination of two box Schubert configurations" in previous observation \cite{Yang:2022gko}. To be more precise, let us focus on four intersections on the {\it e.g.} green circle, which are from the problems ${\color{lightyellow} \mathrm{L}_{3451}}$ and ${\color{lightblue} \mathrm{L}_{3561}}$. Following the Schubert problems,  four intersections correspond to four solutions \eqref{eq:fourso}. Two solutions from the same Schubert configuration ({\it i.e.}, $l_1$ and $l_2$) pinching stands for Landau locus of the problem itself, while pinching of two solutions from different problems ({\it i.e.}, $l_1$ and $l_3$) stands for ${\color{lightred} \mathrm{L}_{34561}}$. Cross-ratios from the four solutions are dual conformal invariants, 
\[\frac{(l_1\cdot l_2)(l_3\cdot l_4)}{(l_1\cdot l_3)(l_2\cdot l_4)}=\left(\frac{\langle1346\rangle\langle2356\rangle}{\langle1356\rangle\langle2346\rangle}\right)^2,\ \frac{(l_1\cdot l_4)(l_2\cdot l_3)}{(l_1\cdot l_3)(l_2\cdot l_4)}=\left(\frac{\langle1236\rangle\langle3456\rangle}{\langle1356\rangle\langle2346\rangle}\right)^2,\]
which account for the letters finally. This also yields an $A_1$ configuration, after projecting solutions $(AB)$ by their intersections on any shared line, {\it i.e.} line $(61)$. We will see this construction by an $A_3$ example in the following. So we summarize our first and most important conclusion in this note.  
\begin{tcolorbox}
  Landau loci always arise from pinching of Schubert solutions, either from an individual or two Schubert configurations. 
  Cross-ratios from solutions account for symbol letters.  
\end{tcolorbox}

For now, we have only considered one subsector of the double-box, that is, Fig.~\ref{fig:6pdbsub_bt} with $\langle CD12\rangle$ shrunk. Considering all the box-triangle subsectors with other propagators shrunk, we will see there are actually two more intersection points on each circle displayed in Fig.~\ref{fig:solution2}, due to the condition $\langle AB12\rangle=0 (=\langle AB23\rangle=\langle AB45\rangle)$, resulting $A_3$ configurations finally. However, here is one more direct way to see this $A_3$, which is from the top integral itself. 

\begin{figure}
    \centering
    \includegraphics[width=0.35\linewidth]{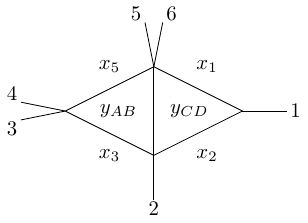}
    \caption{Another double-triangle subtopology of the double-box.}
    \label{fig:6pdbsub_dt1}
\end{figure}

\begin{figure}[htbp]
    \centering
    \includegraphics[width=0.5\textwidth]{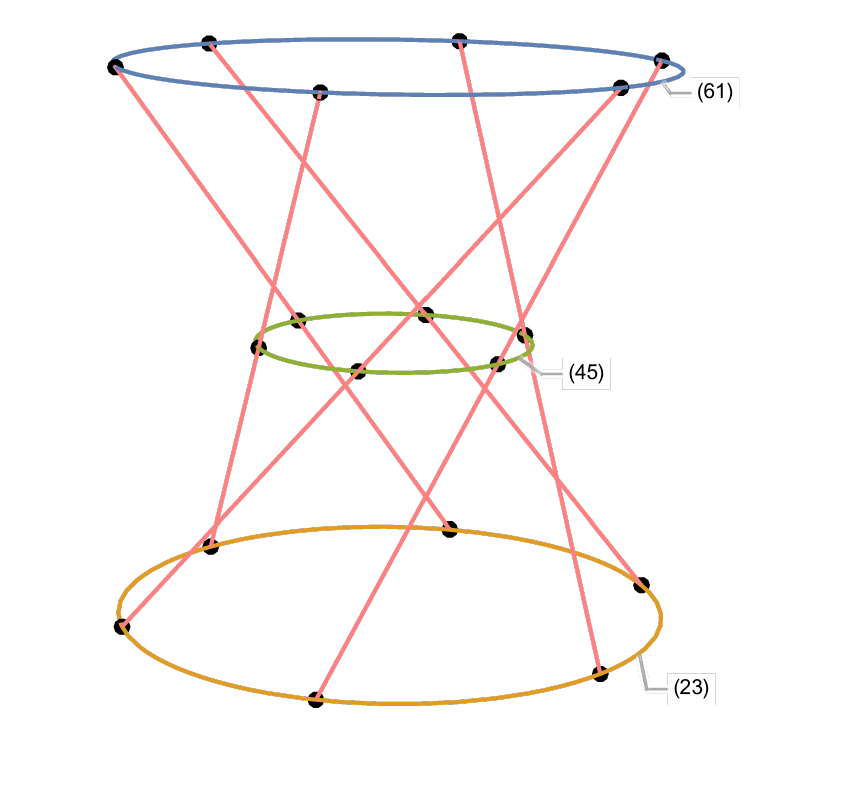}
    \caption{The cross-ratios of the six pink lines can be mapped to the cross-ratios between points on the line $(45)$. In fact we can choose any line on the conic $Q(Z)=\langle Z(45)(61)(23)\rangle$, and $(61)$ and $(23)$ are other two special choices. This suggests a proper map between cross-ratio of lines and cross-ratio of points since the mapping $SL(4)$ is angle-preserving and won't change the shape formed by these lines. So we can define brackets $[ab]$ as in Fig.~\ref{fig:A3} and the cross-ratio between lines will be the square of cross-ratio between points on line $(45)$ or $(61)$, $(23)$. }
    \label{fig:ToMT}
\end{figure}

\begin{figure}[htbp]
    \centering
    \includegraphics[width=\textwidth]{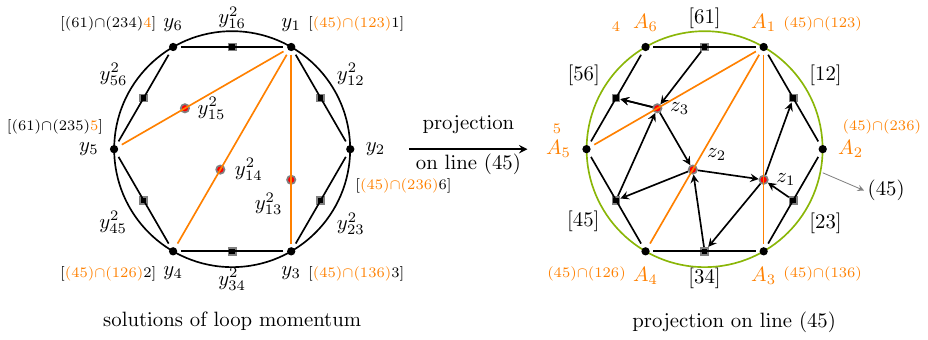}
    \caption{The cross-ratios in momentum space (the left diagram) can capture the singularities when the loop solutions are pinched. The orange color in the expression of lines indicates the part intersecting $(45)$. These cross-ratios are equivalent to the cross-ratios of points after the projection on line $(45)$ (the green circle in the right diagram) in momentum twistor space. Then the $A_3$ cluster algebra structure emerges naturally as the triangulation of a hexagon. We have also drawn the corresponding quiver in the right diagram where red dots are mutable variables and black square blocks are frozen variables. $y_{i}$ is the solutions of loop momentum. $[ab]$ is an abbreviation for $\langle Z_{a}Z_{b}I_{\infty}\rangle$ which is the four bracket between two coordinates with the reference line which we take to be $I_{\infty}$. This projection holds because the cross-ratios between lines in momentum twistor space can be mapped into the cross-ratios between points on the common line $(45)$ which intersects all the lines. This is explained with more details in Fig.~\ref{fig:ToMT}.}
    \label{fig:A3}
\end{figure}

For the top sector of this double-box. The pinch condition for $y_{CD}$ will give us the following constraints (details for the derivation of pinch condition can be found in App.~\ref{app:triangle}.):


\begin{equation}
    \langle AB61\rangle=\langle AB12\rangle=\langle AB56\rangle=0=\langle1256\rangle.
\end{equation}
Combining this with the cut condition for $y_{AB}$ will give us:
\begin{equation}\label{eq:sixconstraints}
    \langle AB23\rangle=\langle AB34\rangle=\langle AB45\rangle=\langle AB61\rangle=\langle AB12\rangle=\langle AB56\rangle=0.
\end{equation}
This is the most constrained condition for $y_{AB}$ in this problem which states the pinch of solutions coming from three different problems\footnote{We should emphasize that although now we have $6$ conditions for $y_{AB}$, at $D=4$ they still result in $\mathbf{Gram}_5$ singularities, since any four conditions are enough to determine $AB$. Plugging the solution in any one of the rest two conditions, we still get $\mathbf{Gram}_5$ singularities. Practically, this constraint is first decomposed into one five-condition constraint and one four-condition constraint. Then the five-condition constraint is further decomposed. All kinds of decomposition give us finally the pinch of three different problems.}. 
For instance, sticking to solutions on the green circle again determined by $\langle AB61\rangle=\langle AB23\rangle=\langle AB45\rangle =0$, three pairs of intersections will appear on that circle, corresponding to three problems: ${\color{lightyellow} \mathrm{L}_{3451}}$ in \eqref{eq:box1}, ${\color{lightblue} \mathrm{L}_{3561}}$ in \eqref{eq:box3} and ${\color{teal} \mathrm{L}_{3512}}$ defined as 
\begin{equation}\label{eq:box5}
    {\color{teal} \mathrm{L}_{3512}}: \langle AB23\rangle=\langle AB45\rangle=\langle AB61\rangle=\langle AB12\rangle=0.
\end{equation}
which corresponds to the double triangle in Fig.~\ref{fig:6pdbsub_dt1}. The solutions of the three problems are drawn in Fig.~\ref{fig:A3}, they are organized in positive order and two points opposite to each other come from the same problem. Here we list them together again for clarity,
\begin{equation}\label{eq:sixsolutions}
    \begin{aligned}
        &{\color{lightyellow} \mathrm{L}_{3451}}: \quad [(61)\cap(234)4], \, [(61)\cap(345)3]; \\
        &{\color{lightblue} \mathrm{L}_{3561}}: \quad [(23)\cap(456)6], \, [(23)\cap(561)5]; \\
        &{\color{teal} \mathrm{L}_{3512}}: \quad [(45)\cap(126)2], \, [(45)\cap(123)1].
    \end{aligned}
\end{equation}
Now let us see this $A_3$ cluster structure by projecting the solutions. This is shown in Fig.~\ref{fig:ToMT} and Fig.~\ref{fig:A3}\footnote{Note that we adopt alternative but equivalent representations for six solutions \eqref{eq:sixsolutions} in Fig.\ref{fig:A3} to emphasize their intersection point on $(45)$. For instance $[(61)\cap(235)5]=[(23)\cap(561)5]$, {\it etc.}}.
Let us explain in more details. First, we consider the one-dimensional solution space of $y_{AB}$ solved by $\langle AB61\rangle=\langle AB23\rangle=\langle AB45\rangle=0$, where the pinches happen. 
Second, we project the solutions in bitwistors to momentum twistors (intersection points) on any external lines and consider their cross-ratios.  The cross-ratios of lines are just the square of cross-ratios of intersection points \cite{He:2023umf}. This is shown in Fig.~\ref{fig:ToMT}. The distance between two points $Z_{a}$ and $Z_{b}$ is defined by $[ab]\equiv\langle Z_{a}Z_{b}I_{\infty}\rangle$. 
So these cross-ratios are naturally defined in Grassmannian $G(2,n)$ where $n=6$ in this example\footnote{The dependence on $I_{\infty}$ will cancel in cross-ratios so we can also choose some arbitrary reference line like $(78)$ in this example.}, and an $A_{3}$ cluster structure appears in this problem\footnote{When $n=6$, we can also see this $A_3$ from cross-ratios of lines directly, since $G(4,6)\simeq G(2,6)$. However, in general $n$, we can only have an $A_{n}$ cluster structure after projecting onto the momentum twistor intersections.}.

We finally provide some details about how to calculate the alphabet from this $A_3$ cluster structure by Fig.~\ref{fig:A3}. We can choose any one of the three lines $(61)$, $(45)$ and $(23)$ to get the $A_3$ letters. For instance, sticking to line $(45)$, the intersections are shown as the right of Fig.~\ref{fig:A3}. The cross-ratios between points on the line  are the $\mathcal{X}$ coordinates as defined in \cite{Fock:2003xxy,Chicherin:2020umh} of corresponding $A_{3}$. The first three of them are defined by the initial quiver depicted in above diagram:
\begin{equation}
    z_{1}=\frac{\langle 1245\rangle\langle 3456\rangle}{\langle 1456\rangle\langle 2345\rangle}, \, z_{2}=\frac{\langle 1235\rangle\langle 1456\rangle}{\langle 1256\rangle\langle 1345\rangle}, \,
    z_{2}=-\frac{\langle 1234\rangle\langle 1256\rangle}{\langle 1236\rangle\langle 1245\rangle}.
\end{equation}
Performing all the mutations which are finite in $A_{3}$, there are nine multiplicative independent polynomials
\begin{equation}\label{eq:A3letter}
    \{z_1,\, z_2,\, z_3,\, 1{+}z_1,\, 1{+}z_2,\, 1{+}z_3,\, 1{+}z_2{+}z_1z_2,\, 1{+}z_3{+}z_2z_3,\, 1{+}z_3{+}z_2z_3{+}z_1z_2z_3\}
\end{equation}
which are exactly multiplicative combinations of the $6$-point $A_3$ letters \eqref{6ptalphabet} discussed in \cite{Dixon:2011pw,Drummond:2014ffa}, {\it etc.}.

We end this section with some remarks. Firstly, as we have mentioned, box-triangle topologies always yield $5$-point Gram determinant singularities, which result in combinations of two Schubert configurations and $A_1$ configurations after geometrizing. When more than one box-triangle sub-topologies are involved in a top integral, we have an $A_{2k{-}1}$ configurations finally from $k$ different box-triangles with a common box-bubble or double-triangle (equivalent to a one-loop box) sub-topology being shared. This is exactly the generation of previous observation \cite{Morales:2022csr} that we should consider all possible combinations for four-mass boxes sharing three external points, because of the box-triangle, box-bubble and double-triangle sub-topologies from $12$-point double-box integral (Fig.\ref{fig:12pdb}), and letters from those $A_3$ exactly read odd letters \eqref{odds12} as well as Gram determinants.

\begin{figure}
    \centering
    \includegraphics[width=0.35\linewidth]{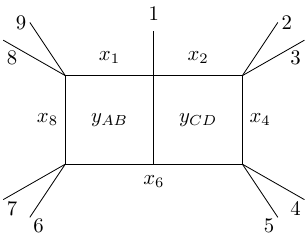}
    \caption{A 9-point double-box integral.}
    \label{fig:9pdoublebox}
\end{figure}

\begin{figure}
    \centering
    \includegraphics[width=0.5\linewidth]{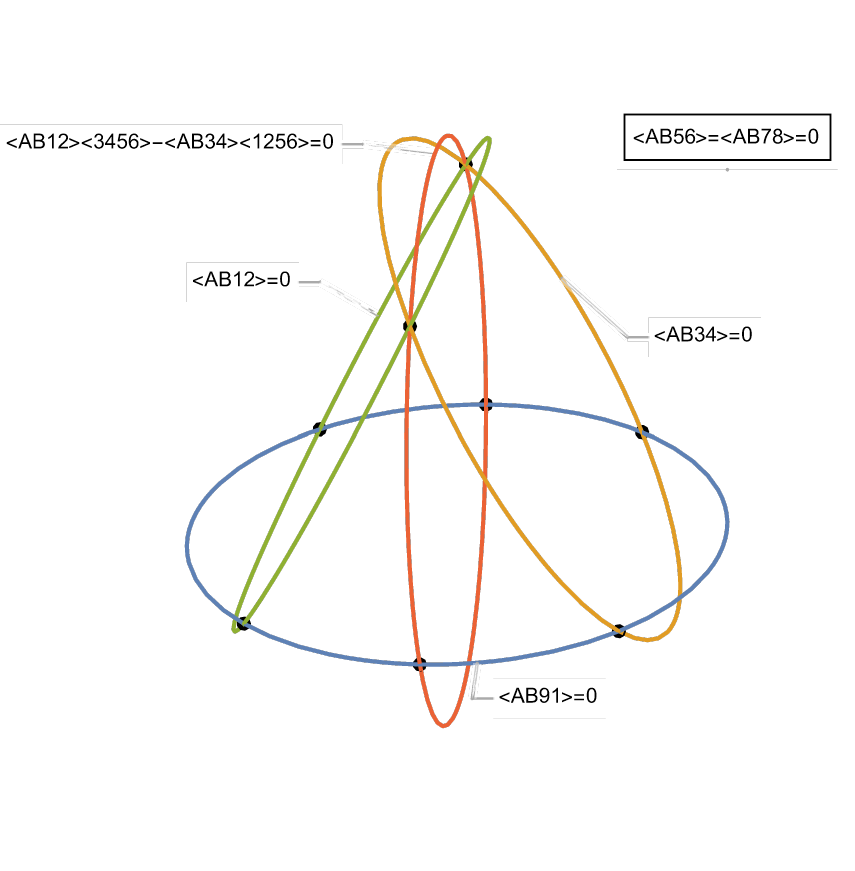}
    \caption{The solutions of $y_{AB}$ in the space under constraint $\langle AB56\rangle=\langle AB78\rangle=0$. We will see there is an additional circle $\langle AB12\rangle\langle 3456\rangle-\langle AB34\rangle\langle 1256\rangle=0$ which is contributed by top sector of double-box in Fig.~\ref{fig:9pdoublebox}.}
    \label{fig:solution3}
\end{figure}

Secondly, in general examples, constraints for kinematics like \eqref{eq:fiveconstraints} from box-triangles and combination of box problems can both be generalized to more general and complicated conditions. Once a possible combination of two sub$^l$-leading Landau equations yield a constraint condition from another sub$^k$-leading sub-topology, we can view the constraint as a combination of these two Schubert configurations and consider the Schubert construction.   As an illustration, we take another more nontrivial example of the 9-point double-box integral, which is one of the sub-topology for 2-loop $\mathrm{N}^2\mathrm{MHV}$ amplitude, and is depicted as in Fig.~\ref{fig:9pdoublebox}.
If we firstly solve equations for $y_{CD}$ and then go to $y_{AB}$, the pinch condition for $y_{CD}$ reads
\begin{equation}
    \langle AB12\rangle\langle3456\rangle-\langle AB34\rangle\langle1256\rangle=0.
\end{equation}
Combining this with the cut conditions for $y_{AB}$, we will have
\begin{equation}
    {{\color{orange}\mathrm{L}_{681\{24\}}}}: \langle AB56\rangle=\langle AB78\rangle=\langle AB91\rangle=\langle AB12\rangle\langle3456\rangle{-}\langle AB34\rangle\langle1256\rangle=0.
\end{equation}
Now for one subsector of this double-box by shrinking $(y_{CD}-x_6)^2$, after pinching $y_{CD}$, we have the following five constraints just like \eqref{eq:fiveconstraints}:
\begin{equation}
    {{\color{lightred}\mathrm{L}_{68124}}}: \langle AB56\rangle=\langle AB78\rangle=\langle AB91\rangle=\langle AB12\rangle=\langle AB34\rangle=0.
\end{equation}
We can see that combinations of ${{\color{orange}\mathrm{L}_{681\{24\}}}}$ with ${{\color{lightgreen}\mathrm{L}_{6812}}}$ or ${{\color{lightyellow}\mathrm{L}_{6814}}}$ defined as
\begin{equation}
    \begin{aligned}
        {{\color{lightgreen}\mathrm{L}_{6812}}}:& \,\langle AB56\rangle=\langle AB78\rangle=\langle AB91\rangle=\langle AB12\rangle=0, \\
        {{\color{lightyellow}\mathrm{L}_{6814}}}:& \,\langle AB56\rangle=\langle AB78\rangle=\langle AB91\rangle=\langle AB34\rangle=0,
    \end{aligned}
\end{equation}
can give ${{\color{lightred}\mathrm{L}_{68124}}}$ too.  Geometrically, this just modifies the Fig.~\ref{fig:solution2} to be like Fig.~\ref{fig:solution3}.
The top sector will contribute another pair of black dots in the subspace determined by $\langle AB56\rangle=\langle AB78\rangle=\langle AB91\rangle=0$ (the blue circle), and we have some new $A_3$ configurations instead. The discussion of pinches between points in the same one-dimensional space will be the same as before, and the cross-ratios give us letters again, which, in this case, include the non-trivial odd letters mixing two different square roots from ${\color{orange}\mathrm{L}_{681\{24\}}}$ and ${\color{lightyellow}\mathrm{L}_{6814}}$. To illustrate this, let us list the six points on the blue circle labeled by their origin:
\begin{equation}
    \begin{aligned}
        {\color{orange}\mathrm{L}_{681\{24\}}}:&\,\,l_{1}^{\pm}=(197)\cap(567)+\alpha_{\pm}[(197)\cap(568)+(198)\cap(567)]+\alpha_{\pm}^2(198)\cap(568) \\{\color{lightyellow}\mathrm{L}_{6814}}:&\,\,l_{2}^{\pm}=(197)\cap(567)+\beta_{\pm}[(197)\cap(568)+(198)\cap(567)]+\beta_{\pm}^2(198)\cap(568)\\
        {{\color{lightgreen}\mathrm{L}_{6812}}}:&\,\, l_{3}^{+}=(78)\cap(912)(56)\cap(912), \, l_{3}^{-}=(178)\cap(156)
    \end{aligned}
\end{equation}
where $\alpha_{\pm}$ and $\beta_{\pm}$ involve two different square roots
\begin{equation}
\begin{aligned}
    \alpha_{\pm}&=\frac{\langle(197)\cap(568)34\rangle\langle1256\rangle-\langle(197)\cap(568)12\rangle\langle3456\rangle+(7\leftrightarrow 8)\pm \sqrt{\Delta_{6}}}{2(\langle(198)\cap(568)12\rangle\langle3456\rangle-\langle(198)\cap(568)34\rangle\langle1256\rangle)},\\
    \beta_{\pm}&=\frac{-\langle(197)\cap(568)34\rangle-\langle(198)\cap(567)34\rangle\pm \sqrt{\mathbf{Gram}_{6,8,1,4}}}{2\langle(198)\cap(568)34\rangle},\\
\end{aligned}
\end{equation}
where $\Delta_{6}$ is the leading singularity of this top sector,
\begin{equation}
    \begin{aligned}
        \Delta_{6}=&(x_{28}^2x_{16}^{2}x_{46}^2-x_{26}^2x_{16}^{2}x_{48}^2+x_{26}^2x_{14}^2x_{68}^2)^2-4x_{24}^2x_{26}^2x_{16}^2x_{18}^2x_{46}^2x_{68}^2 \\
        =&(\langle1278\rangle\langle1569\rangle\langle3456\rangle-\langle1256\rangle\langle1569\rangle\langle3478\rangle+\langle1256\rangle\langle1349\rangle\langle5678\rangle)^2 \\&-4\langle1234\rangle\langle1256\rangle\langle1569\rangle\langle1789\rangle\langle3456\rangle\langle5678\rangle.
    \end{aligned}
\end{equation}
Now the pinches between solutions of ${\color{orange}\mathrm{L}_{681\{24\}}}$ and ${\color{lightyellow}\mathrm{L}_{6814}}$ have very simple forms:
\begin{equation}
\begin{aligned}
    &(l_1^{\pm}\cdot l_2^{\pm})=-\langle 5691\rangle\langle 7891\rangle\langle 5678\rangle(\alpha_{\pm}-\beta_{\pm})^2, \\&(l_1^{+}\cdot l_{1}^{-})=\frac{-\langle 5691\rangle\langle 7891\rangle\langle 5678\rangle\Delta_{6}}{(\langle(198)\cap(568)12\rangle\langle3456\rangle-\langle(198)\cap(568)34\rangle\langle1256\rangle)^2}, \\
    &(l_2^{+}\cdot l_2^{-})=\frac{-\langle 5691\rangle\langle 7891\rangle\langle 5678\rangle\mathbf{Gram}_{6,8,1,4}}{(\langle(198)\cap(568)34\rangle)^2}.
\end{aligned}
\end{equation}
The cross ratio between $l_{1}^{\pm}$ and $l_{2}^{\pm}$ (the first line) will give us the non-trivial odd letter
\begin{equation}
    \begin{aligned}
        \frac{(l_{1}^{+}\cdot l_{2}^{+})(l_{1}^{-}\cdot l_{2}^{-})}{(l_{1}^{+}\cdot l_{2}^{-})(l_{1}^{-}\cdot l_{2}^{+})}=\left[\frac{(\alpha_{+}-\beta_{+})(\alpha_{-}-\beta_{-})}{(\alpha_{+}-\beta_{-})(\alpha_{-}-\beta_{+})}\right]^2=\left[\frac{A+\sqrt{\Delta_{6}\mathbf{Gram}_{6,8,1,4}}}{A-\sqrt{\Delta_{6}\mathbf{Gram}_{6,8,1,4}}}\right]^2
    \end{aligned}
\end{equation}
where
\begin{equation}
    \begin{aligned}
        A=&-x_{26}^2 x_{48}^4 x_{16}^4+x_{28}^2 x_{46}^2 x_{48}^2 x_{16}^4-x_{18}^2 x_{28}^2 x_{46}^4 x_{16}^2+x_{18}^2 x_{26}^2 x_{46}^2 x_{48}^2 x_{16}^2+x_{14}^2 x_{18}^2
   x_{26}^2 x_{46}^2 x_{68}^2 \\
        &+2 x_{18}^2 x_{24}^2
   x_{46}^2 x_{68}^2 x_{16}^2-x_{14}^2 x_{28}^2 x_{46}^2 x_{68}^2 x_{16}^2+2 x_{14}^2 x_{26}^2 x_{48}^2 x_{68}^2 x_{16}^2-x_{14}^4 x_{26}^2 x_{68}^4.
    \end{aligned}
\end{equation}
Explicit computation in \cite{Wilhelm:2022wow} showed that this letter reads one of the 4 last entries of the 9-point integral, and all other odd letters for this integral, as discussed in \cite{Yang:2022gko}, can be constructed in a similar way.


\subsection{An example of elliptic Feynman integrals}\label{ellipticsec}
\begin{figure}[htbp]
    \centering
    \includegraphics[width=0.3\linewidth]{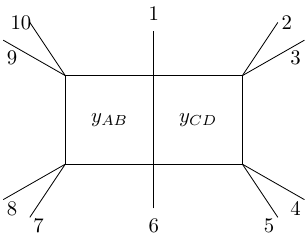}
    \caption{The 10-point elliptic ladder diagram.}
    \label{fig:ellipticladder}
\end{figure}

When studying Landau equations and Landau loci, one does not need to distinguish between cases when Feynman integrals evaluate to MPL functions (where the notion of ``symbol" clearly applies) and more complicated cases when they evaluate to {\it e.g.} elliptic MPL~\cite{Broedel:2018iwv,Broedel:2019hyg,Kristensson:2021ani,Wilhelm:2022wow} or functions that are even less understood~\cite{Bourjaily:2019hmc}. This is because Landau analysis only concerns locations of singularities, but such a distinction becomes natural when we consider the one-dimensional solution space and study Schubert analysis. Let us take the 10-point elliptic double-box depicted in Fig.~\ref{fig:ellipticladder} as an example and explain how our method applies to such cases. 

\begin{figure}[htbp]
    \centering
    \includegraphics[width=0.3\linewidth]{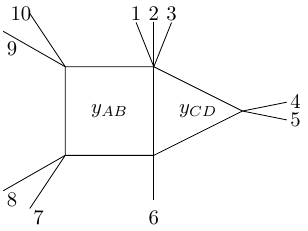}
    \caption{One box-triangle subsector of elliptic double-box in Fig.~\ref{fig:ellipticladder}.}
    \label{fig:ellipticbt}
\end{figure}

For the top sector, the pinch condition for $y_{CD}$ will give us
\begin{equation}\label{eq:10pdbpinchAB}
        \left(\begin{array}{cccc}
            0 & \langle1234\rangle & \langle1256\rangle & \langle AB12\rangle \\
            \langle1234\rangle & 0 & \langle3456\rangle & \langle AB34\rangle \\
            \langle1256\rangle & \langle3456\rangle & 0 & \langle AB56\rangle \\
            \langle AB12\rangle & \langle AB34\rangle & \langle AB56\rangle & 0
        \end{array}\right)
        \left(\begin{array}{c}
            \beta_1 \\ \beta_2 \\ \beta_3 \\ \gamma
        \end{array}\right)=0 \, .
\end{equation}
Again, the requirement that this system has non-trivial solutions that $\beta_{1},\beta_{2},\beta_{3},\gamma\neq 0$ gives us the following constraint
\begin{equation}\label{eq:ellipticpinch}
    \left(\langle1234\rangle\langle AB56 \rangle\!-\!\langle1256\rangle\langle AB34 \rangle\!-\!\langle3456\rangle\langle AB12 \rangle\right)^2\!-\!4\langle1256\rangle\langle AB34 \rangle\langle3456\rangle\langle AB12 \rangle=0.
\end{equation}
Combining it with the three cut conditions for $y_{AB}$:
\begin{equation}\label{eq:ellipticcut}
    \langle AB10\ 1\rangle=\langle AB89\rangle=\langle AB67\rangle=0 ,
\end{equation}
This will actually give four nontrivial solutions of $y_{AB}$ because \eqref{eq:ellipticpinch} will be a quartic polynomial of the remaining parameter of the one-dimensional solution space determined by \eqref{eq:ellipticcut}. We abbreviate \eqref{eq:ellipticpinch} and \eqref{eq:ellipticcut} as ${{\color{orange}\mathrm{L}_{e791}}}$. Now let us search the whole subsectors to find other solutions of $y_{AB}$ which share the same one-dimensional space determined by \eqref{eq:ellipticcut}. They come from three box-bubble diagrams:
\begin{equation}\label{eq:ellipticboxes}
    \begin{aligned}
        {{\color{lightgreen}\mathrm{L}_{6791}}}:& \,\langle AB10\ 1\rangle=\langle AB89\rangle=\langle AB67\rangle=\langle AB56\rangle=0, \\
        {{\color{lightyellow}\mathrm{L}_{4791}}}:& \,\langle AB10\ 1\rangle=\langle AB89\rangle=\langle AB67\rangle=\langle AB34\rangle=0, \\
        {{\color{lightblue}\mathrm{L}_{2791}}}:& \,\langle AB10\ 1\rangle=\langle AB89\rangle=\langle AB67\rangle=\langle AB12\rangle=0 .
    \end{aligned}
\end{equation}
However, to determine which solutions can be pinched, we must also find the corresponding Landau equations for the pinches. These usually comes from diagrams with triangle structures. Let us take one such box-triangle depicted in Fig.~\ref{fig:ellipticbt} as an example.
The Landau equations for $y_{AB}$ will be
\begin{equation}\label{eq:ellipticfiveconstraints}
\begin{aligned}
    {{\color{lightred}\mathrm{L}_{46791}}}:\,\langle AB10\ 1\rangle=\langle AB89\rangle=\langle AB67\rangle=\langle AB34\rangle=\langle AB56\rangle=0.
\end{aligned}
\end{equation}
Actually, there is also a kinematic constraint for this box-triangle diagram, $\langle3456\rangle=0$. That is, \eqref{eq:ellipticfiveconstraints} should hold under the condition $\langle3456\rangle=0$. This kinematic condition is either absent or not relevant in the former simple examples. However, in the current case, it will play an important role as we will see.

First, we can find that the combination of ${{\color{lightgreen}\mathrm{L}_{6791}}}$ and ${{\color{lightyellow}\mathrm{L}_{4791}}}$ will give ${{\color{lightred}\mathrm{L}_{46791}}}$ no matter whether $\langle3456\rangle=0$ holds or not. This means we can always pinch solutions from these two problems. In the same time, we know the simultaneous pinch of three pairs of solutions in \eqref{eq:ellipticboxes} is forbidden because there is no such Landau equation
\begin{equation}
    \langle AB10\ 1\rangle=\langle AB89\rangle=\langle AB67\rangle=\langle AB34\rangle=\langle AB56\rangle=\langle AB12\rangle=0
\end{equation}
that all the six are 0 in this integral system. Then what about the pinch between ${{\color{orange}\mathrm{L}_{e791}}}$ and ${{\color{lightgreen}\mathrm{L}_{6791}}}$? At first sight, there are no corresponding Landau equations in this system too since the combination of ${{\color{orange}\mathrm{L}_{e791}}}$ and ${{\color{lightgreen}\mathrm{L}_{6791}}}$ gives us
\begin{equation}\label{eq:ellipticcombine}
    \begin{aligned}
        &\langle AB10\ 1\rangle=\langle AB89\rangle=\langle AB67\rangle=\langle AB56\rangle, \\
        &\langle1256\rangle\langle AB34\rangle-\langle3456\rangle\langle AB12\rangle=0 .
    \end{aligned}
\end{equation}
No subsectors of this family can give such Landau equations for $y_{AB}$\footnote{It is interesting to note that these are Landau equations for a super-sector of this family by turning the left box of this integral into a pentagon. We also note that the condition ${{\color{lightgreen}\mathrm{L}_{6791}}}$ reduces the pinch condition \eqref{eq:ellipticpinch} which is related to an ellitpic curve to a perfect square and this gives the second line of \eqref{eq:ellipticcombine}. This can also be interpreted as a pinch between four solutions of the quartic polynomial \eqref{eq:ellipticpinch}.}. However, we can see that when $\langle3456\rangle=0$, the combination of these two gives us ${{\color{lightred}\mathrm{L}_{46791}}}$\footnote{Here we suppose $\langle1256\rangle$ is general.}. To summarize, we can pinch solutions from these two problems under the condition $\langle3456\rangle=0$ while for general $\langle3456\rangle$ this pinch will give us new Landau singularities that seems not existing in this system.
For the same reason, we can not pinch ${{\color{orange}\mathrm{L}_{e791}}}$, ${{\color{lightgreen}\mathrm{L}_{6791}}}$ and ${{\color{lightyellow}\mathrm{L}_{4791}}}$ simultaneously with general $\langle3456\rangle$, however, this can be done when $\langle3456\rangle=0$. This indicates that we should somehow generalize this pinch picture for more involved cases.

\section{Schubert analysis from Landau diagrams in $\mathcal{N}=4$ SYM}

In this section, we move on to applying this Landau-based Schubert analysis for physical quantities such as scattering amplitudes and form factors in planar $\mathcal{N}=4$ SYM. Note that as already noted in~\cite{Dennen:2015bet}, although the amplitude can be expressed in terms of DCI integrals with unit leading singularities ({\it c.f.} \cite{Arkani-Hamed:2010zjl} where {\it e.g.} the two-loop MHV and NMHV amplitudes can be expressed in terms of double-pentagon integrals with chiral numerators), there are way more singularities as obtained from the Landau analysis for these integrals than singularities in the final amplitudes, since huge cancellations happen when we sum over these integrals. In \cite{Prlina:2017azl,Prlina:2017tvx}, the authors alternatively proposed ``Landau diagrams" for directly computing singularities of scattering amplitudes (as opposed to those individual integrals), which nicely follow from boundary structures of the amplituhedra~\cite{Arkani-Hamed:2013jha}. Here we will follow these Landau diagrams for amplitudes and apply Schubert analysis, which amounts to ``uplifting" Landau singularities to symbol letters. As we will see, our method will not only produce the correct symbol alphabet for two-loop MHV and NMHV amplitudes, but also reveal the origin of cluster algebraic structures therein.

Let us first give a brief review of amplituhedra and Landau diagrams for scattering amplitudes in $\mathcal{N}=4$ SYM theory, which is the starting point for our computations. Since directly applying Landau analysis to individual integrals leads to spurious singularities, we have to distinguish the physical loci from the unphysical ones. An important idea is that all physical singularities and corresponding solutions for loop momenta in \eqref{cut} and \eqref{pinch} should accord with geometrical boundaries of the amplituhedron~\cite{Arkani-Hamed:2013jha}, {\it i.e.} solutions for loop momenta should always locate in interior of the geometry. 

Roughly speaking, the amplituhedron $\mathcal{A}_{n,k,L}$ for $n$-point $L$-loop N$^k$MHV amplitude is a collection of pairs $(Y,\mathcal{L})$, consisting of $Y\in Gr(k,k{+}4)$ and $L$ lines $\mathcal{L}_1,\cdots, \mathcal{L}_{L}\in Gr(2,k{+}4)$ in complement of $Y$, and its interior is described as (following the sign flip definition in~\cite{Arkani-Hamed:2017vfh}): (1) external kinematics data are restricted in the principle domain $\mathcal{D}_{n,k}$, which is defined as $\langle ii{+}1jj{+}1\rangle>0$ and $\langle 123k\rangle$ for $k=4,\cdots,n$ having $k$ sign flips; (2) $\langle Y\mathcal{L}_i kk{+}1\rangle>0$ ($(-1)^{k{+}1}\langle Y\mathcal{L}_i n1\rangle>0$) for all $i$ and $k$; (3) $\langle Y\mathcal{L}_i 1k\rangle>0$ has $k{+}2$ sign flip; (4) $\langle Y\mathcal{L}_i\mathcal{L}_j\rangle>0$ for all pairs $i$ and $j$. Canonical form of the geometry yields integrand for scattering amplitudes. Readers can also refer to \cite{Arkani-Hamed:2013jha,Arkani-Hamed:2017vfh} for more details.

\begin{figure}
    \centering
      \includegraphics[width=0.3\linewidth]{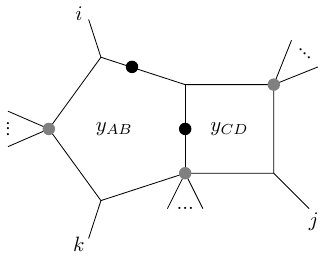}
    \caption{2-loop MHV Landau diagram}
    \label{fig:2LMHV}
\end{figure}

Such a definition therefore offers an important criterion for all physical solutions of loop momenta from Landau equation. For each special $n,k,L$ case,  We should only include those solutions $\mathcal{L}_i$ satisfying the sign flip definition of $\mathcal{A}_{n,k,L}$, while get rid of other solutions outside the geometry. In~\cite{Prlina:2017azl,Prlina:2017tvx}, the authors classified all physical boundaries for $n$-point $L=1$ and $2$ N$^k$MHV amplituhedra. Consequentially, all physical maximal cut solutions should correspond to these physical boundaries, indicating that they should be Landau equation solutions for several special {\it Landau diagrams}. Computing singularities for the amplitudes then boils down to analyzing physical singularities for these diagrams, and so does the symbol letter via Schubert analysis, as we will see shortly. The simplest MHV case suffices to illustrate our idea, which can be applied to N$^k$MHV amplitudes. Following the analysis in~\cite{Prlina:2017azl,Prlina:2017tvx}, we only need to consider the diagram shown in Fig.\ref{fig:2LMHV} for $i<j<k$.

\subsection{From Landau to Schubert for two-loop MHV amplitudes}
\begin{figure}[htbp]
    \centering
    \includegraphics[width=0.8\linewidth]{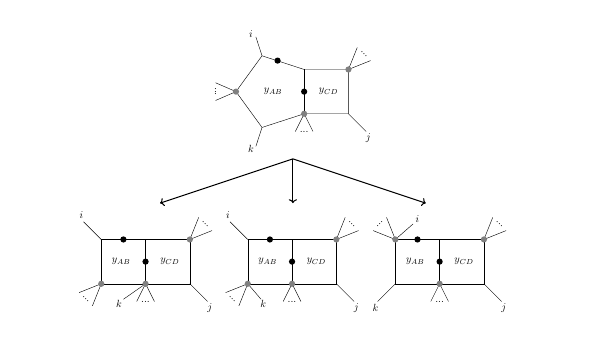}
    \caption{Landau diagrams for 2-loop MHV amplitudes I}
    \label{fig:MHV1}
\end{figure}

Let us firstly revisit the Landau singularities for two-loop amplitudes and define all Schubert configurations associated with the diagram in Fig.~\ref{fig:2LMHV}, on which we perform the Schubert analysis. We list all its sub-topologies with at most two propagators shrunk in Fig.~\ref{fig:MHV1} and Fig.~\ref{fig:MHV2} (diagrams with more propagators shrunk, like box-bubbles or double-triangles, will be equivalent to one-loop problems, as we have mentioned), and we apply to them the Landau/Schubert analysis. Note that we don't arrange the diagrams by their topologies, but by the constraints that they impose on $y_{AB}$ after solving Landau equation for $y_{CD}$ first. 
Explicit calculation of their Landau loci can be found in our \verb|Mathematica| notebook in the ancillary file:

\begin{itemize}
    \item For the top diagram in Fig.~\ref{fig:MHV1}, constraints imposed on $(AB)$ are
    \begin{equation}\label{eq:MHVtopConstraints}
        \begin{aligned}
            &\text{cuts: } \langle AB k-1 k \rangle=\langle AB k k+1 \rangle=\langle AB i-1 i \rangle=\langle AB i i+1 \rangle = 0,\\
            &\text{pinch: } \langle ABj{-}1j\rangle\langle ii{+}1jj{+}1\rangle-\langle ABjj{+}1\rangle\langle ii{+}1j{-}1j\rangle=\langle AB(ii{+}1j)\cap(\bar{j})\rangle=0.
        \end{aligned}
    \end{equation}
    Together with constraints on $(CD)$, its straightforward to get the Landau loci of this diagram, which is the 5-point Gram determinant formed by 
    \[\{(i{-}1i),(ii{+}1),(k{-}1k),(kk{+}1),(ii{+}1j)\cap(\bar{j})\}\]
    Explicit computation shows that it reads (the square-root of)
    \[
\frac{(\mathbf{Gram}_{i{+}1,j,j{+}1,k,k{+}1})^2\mathbf{Gram}_{i,i{+}1,j,j{+}1}\mathbf{Gram}_{i,i{+}1,k,k{+}1}}{\mathbf{Gram}_{j,j{+}1,k,k{+}1}}\]

    
    \item For the second line in Fig.~\ref{fig:MHV1}, by shrinking one propagator on the pentagon side, we have 3 different kinds of constraints by deleting one of the following constraint from Eq.~\eqref{eq:MHVtopConstraints}.
    \begin{equation}
    \langle AB k-1 k \rangle=0, \, \langle AB k k+1 \rangle=0, \, \langle AB i-1 i \rangle=0.\\
    \end{equation}
    Each of these constraints yields a ``box" problems, and corresponding singularities reads the 4-point Gram determinant formed by the four dual points. Explicit computation confirms three singularities correspondingly as
    \[\left\{\langle ii{+}1kk{+}1\rangle^2\mathbf{Gram}_{i,i{+}1,j,j{+}1},\ \langle ii{+}1k{-}1k\rangle^2\mathbf{Gram}_{i,i{+}1,j,j{+}1},\ \frac{(\mathbf{Gram}_{i{+}1,j,j{+}1,k,k{+}1})^2}{4\mathbf{Gram}_{j,j{+}1,k,k{+}1}}\right\}\]
    
    \item The top diagram in Fig.~\ref{fig:MHV2} is obtained by shrinking $\langle CDii{+}1\rangle$. This topology is the most complicated one, which gives the following 6 constraints on $(AB)$:
    \begin{equation}\label{eq:sixConstaints}
        \langle AB i-1 i \rangle = \langle AB i i+1 \rangle = \langle AB j-1 j \rangle = \langle AB j j+1 \rangle = \langle AB k-1 k \rangle = \langle AB k k+1 \rangle = 0.
    \end{equation}
    However, as we have discussed in the last section, these constraints still result in all possible $5$-point Gram determinants by choosing any five conditions from them and computing the locus. It is easy to see that they contribute $\mathbf{Gram}_{a,a{+}1,b,b{+}1,c}$ for all possible $\{a,b,c\}\in[n]$
    
    \item Then for the second line in Fig.~\ref{fig:MHV2}, 5 constraints will be imposed on $(AB)$, yielding $\mathbf{Gram}_{a,a{+}1,b,b{+}1,c}$ exactly, and diagrams put in one box give the same constraints. Each kind of constraints is obtained by deleting one of the constraint in Eq.~\eqref{eq:sixConstaints} except $\langle AB i i+1 \rangle = 0$.
    \item For the third line in Fig.~\ref{fig:MHV2}, they all give different constraints of the box type, yielding 4-point Gram determinant (up to $\mathbf{Gram}_{a,a{+}1,b,c}$) as their singularities, which is easily read off from the corresponding diagrams. 
\end{itemize}


We can therefore summarize all singularities appear in two-loop MHV amplitudes; for $n$-point two-loop amplitudes all possible singularities can only be 4-point and 5-point Gram determinants as the following 
\begin{equation}\label{MHVsingularities}
    \{\langle aa{+}1bb{+}1\rangle,\ \mathbf{Gram}_{a,a{+}1,b,c},\ \mathbf{Gram}_{a,a{+}1,b,b{+}1,c}\}_{a,b,c\in[n]}
\end{equation}
In $\mathcal{A}$-coordinates, these $\mathbf{Gram}_A$ are just
\begin{align}  &\mathbf{Gram}_{a,a{+}1,a{+}2,a{+}3}=\langle a{-}1 a a{+}1 a{+}2\rangle^2\langle a a{+}1 a{+}2 a{+}3\rangle^2\\
&\mathbf{Gram}_{a,a{+}1,a{+}2,b}=\langle a{-}1 a a{+}1 a{+}2\rangle^2\langle a a{+}1 b{-}1 b\rangle^2\\
&\mathbf{Gram}_{a,a{+}1,b,b{+}1}=\langle a{-}1 a a{+}1 b\rangle^2\langle a b{-}1 b b{+}1\rangle^2\\
&\mathbf{Gram}_{a,a{+}1,b,c}=\langle a (a{-}1 a{+}1) (b{-}1b)(c{-}1c)\rangle^2\\
&\mathbf{Gram}_{a,a{+}1,b,b{+}1,c}={-}2\langle a{-}1 a a{+}1 b\rangle\langle a b{-}1 b b{+}1\rangle\langle c{-}1c (a{-}1 a a{+}1)\cap(b{-}1bb{+}1)\rangle\langle c{-}1c ab\rangle
\end{align}
whose factors are exactly all two-loop letters for MHV amplitudes \cite{Caron-Huot:2011zgw}. 

On the Schubert side, we have all one-loop box Schubert configurations up to three-mass, as well as three non-trivial new Schubert configurations from the second line of Fig.\ref{fig:MHV1} as we have mentioned. Now we are ready to construct letters and $A_n$ configurations from these configurations. 

\subsection{Schubert analysis and the alphabet as union of $A_3$}

\paragraph{$\mathbf{n=6}$ and $\mathbf{7}$ } Let us begin with $n=6$ case.
As we have mentioned, all possible singularities are 4- or 5-point Gram determinants \eqref{MHVsingularities}. In 6-point case, these are just all possible Gram determinants 
\[\{\mathbf{Gram}_{i,j,k,l},\mathbf{Gram}_{i,j,k,l,m},\ \ i,j,k,l,m\in\{1,2,3,4,5,6\}\}\]
Rewriting these letters in $\mathcal{A}$-coordinates, all the Grams are converted to $12$ combinations of Pl\"ucker variables as
\begin{equation}\label{6ptrational}
\{\langle i\ i{+}1\ i{+}2\ i{+}3\rangle, \langle i\ i{+}1\ j\ j{+}1\rangle,\ \langle\bar{i}j\rangle\langle i\bar j\rangle\}
\end{equation}
We see that all 15 Pl\"ucker variables have already appeared as factors. 

\begin{figure}[htbp]
    \centering
    \includegraphics[width=0.3\linewidth]{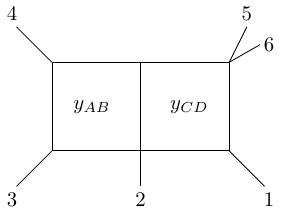}
    \includegraphics[width=0.3\linewidth]{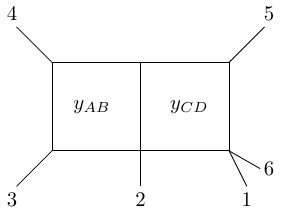}
    \includegraphics[width=0.3\linewidth]{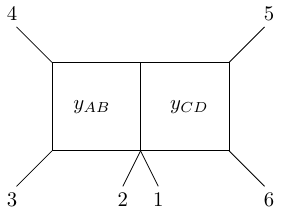}
    \caption{Non-trivial double-box subtopologies of 6-point MHV Landau diagram.}
    \label{fig:6pmhvdb}
\end{figure}

Now let us geometrize cut conditions and letters to Schubert configuration, and see that  $A_3$ cluster algebraic structures appear naturally for these Pl\"ucker variables.  As discussed generally in the last subsection, all individual Schubert configurations we need to look into are three non-trivial new Schubert configurations and its dihedral images from two-loop integrals in Fig.~\ref{fig:6pmhvdb}, together with one-loop boxes configurations from all possible double-triangles/box-bubble sub-topologies.
Here is a very important shortcut for our discussion: the three non-trivial two-loop integrals are always equivalent to one-loop boxes again, if we always choose to firstly solve Landau equation for $y_{AB}$ and then $y_{CD}$, {\it i.e.} pinch condition for $y_{AB}$ in all the three cases always yield $\langle CD34\rangle=0$ for $y_{CD}$. Therefore, we can equivalently only consider all boxes Schubert configurations (up to two-mass in this case due to $n=6$) and their combinations.

Next we go to the $5$-point Gram determinants, where the combinations of boxes arise. All these singularities are derived from configurations like
\begin{equation}\label{Gram51}
\langle AB61\rangle=\langle AB12\rangle=\langle AB23\rangle=\langle AB34\rangle=\langle AB45\rangle=0
\end{equation}
which can be seen originating from the box-triangle subtopology, as well as penta-triangle topologies as we have discussed. Again, since equation \eqref{Gram51} cannot be satisfied unless external kinematics is special, we should regard the five conditions as a combination of two box Schubert configurations sharing three external points. Following a similar discussion as the example in section 2, for any three lines $\{(ii{+}1),(jj{+}1),(kk{+}1)\}$ with $1\leq i,j,k\leq6$, we have triples of boxes. Solutions from any pair of the three boxes pinch yielding $5$-point Gram determinant singularities, while solutions for any one of the three boxes pinch yielding $4$-point Gram. Especially, when we considering three boxes sharing $(61)$, $(23)$, $(45)$, explicit computation shows that we get $A_3$  configuration in momentum twistor space 
with six intersections ordered in positive region of Grassmannian $G(4,6)$ as ({\it e.g.} on the line $(45)$ as Fig.\ref{fig:A3})
\[\{(45)\cap(126),\ (45)\cap(136),\ (45)\cap(236),\ (45)\cap(123),4,5\}\]
and from this $A_3$ we get exactly the $9$ letters in the amplitudes which are defined in \eqref{eq:A3letter}.  One can also consider other triple of boxes to get 5-point Gram singularities. The upshot is that only considering three boxes sharing $\{1,3,5\}$ or $\{2,4,6\}$ give us $A_3$ on external lines, while from other triples we only have degenerated $A_1$ on shared external lines, and they do not produce new letters. This is exactly the generation of $A_3$ cluster algebraic structures for $n=6$ amplitudes.


At $n=7$, all independent Schubert configurations for MHV amplitudes and possible combinations are
\begin{itemize}
    \item All one-loop boxes $\{a,a{+}1,b,c\}$ from double-triangles and box-bubbles. Following the most complicated condition \eqref{eq:sixConstaints}, we can consider all possible combinations of 3 boxes from six dual points $\{i,i{+}1,j,j{+}1,k,k{+}1\}$ sharing three external lines. The most complicated combinations are three boxes sharing $\{(12),(34),(56)\}$ and all cyclic and dihedral images, from which we get $A_3$ configurations as well.
    \item Four different new Schubert configurations from two-loop topologies and their cyclic images. They are
    \begin{align}
        &\langle AB67\rangle=\langle AB71\rangle=\langle AB12\rangle=\langle AB45\rangle\langle1256\rangle-\langle AB56\rangle\langle1245\rangle=0\label{eq:7ptex}\\
        &\langle AB56\rangle=\langle AB71\rangle=\langle AB12\rangle=\langle AB34\rangle\langle1245\rangle-\langle AB45\rangle\langle1234\rangle=0\\
        &\langle AB56\rangle=\langle AB67\rangle=\langle AB12\rangle=\langle AB34\rangle\langle1245\rangle-\langle AB45\rangle\langle1234\rangle=0\\
        &\langle AB67\rangle=\langle AB71\rangle=\langle AB12\rangle=\langle AB34\rangle\langle1245\rangle-\langle AB45\rangle\langle1234\rangle=0
    \end{align}
    Each of them can be combined with two boxes ({\it e.g.} \eqref{eq:7ptex} with boxes $(1,2,5,7)$ and $(1,2,6,7)$), yielding 5-Gram singularities and $A_3$ configurations again.
\end{itemize}
Union all these $A_3$ configurations together, we can get exactly $42$ symbol letters as the following
\begin{align}\label{7ptalphabet2}
\{u_i,1{-}u_i,&1{-}u_i u_{i{+}3},1{-}u_i u_{i{+}3}{-}u_{i{+}1}u_{i{+}5}\}_{i=1,\cdots, 7}\nonumber\\
&\cup\biggl\{y_{i1}{=}\frac{{-}1{+}u_i{-}u_{i{+}1}{+}u_iu_{i{+}1}u_{i{+}4}{+}u_{i{+}2}u_{i{-}1}{+}\Delta_i}{{-}1{+}u_i{-}u_{i{+}1}{+}u_iu_{i{+}1}u_{i{+}4}{+}u_{i{+}2}u_{i{-}1}{-}\Delta_i},y_{i2}{=}y_{i1}(u_i\leftrightarrow u_{i{+}1})\biggr\}_{i=1\cdots7}
\end{align}
where
$u_1=\frac{\langle1234\rangle\langle4571\rangle}{\langle7134\rangle\langle1245\rangle}$ and $\Delta_1{=}\sqrt{(1{-}u_1{-}u_2{+}u_1u_2u_5{-}u_3u_7)^2{-}4 u_1 u_2 u_3 u_7(1-u_5)}$. It remains an interesting question that how these $A_3$ alphabet form $E_6$ cluster algebraic structures \cite{Caron-Huot:2020bkp} finally.

\paragraph{$\mathbf{n\geq8}$ and $A_3$ configuration union for MHV alphabet}
Similar logic applies to $n\geq8$. Note that from $n=8$, not all 4- and 5-point Gram determinants are singularities for MHV amplitudes. Firstly, four-mass boxes and $5$-point Gram containing four-mass box are absent in \eqref{MHVsingularities}, Secondly, beginning with $n=8$, we have $\mathbf{Gram}_{1,2,3,5,7}$, which does not belong to \eqref{MHVsingularities} as well. Therefore, we only need to consider boxes up to three mass, and the allowed combinations for three boxes following \eqref{eq:sixConstaints} always involve six dual points with three-mass-easy hexagon topology (Fig.\ref{fig:2LMHV}). Besides, three two-loop non-trivial topologies in the last line of Fig.\ref{fig:MHV1} yield extra $A_3$ following a similar logic. These configuration, however, offer no new letters other than $A_3$ from boxes. Therefore we can focus on the $A_3$ from box combinations and their union.

In general $n$, the most complicated combinations for boxes, {\it i.e.} combinations of $3$ boxes with $\{(i{-}1 i),(j{-}1 j), (k{-}1k)\}$ shared, are like (any $a{+}1$ in the list can be replaced by $a{-}1$)
\[\{(i,j,k,i{+}1),\ (i,j,k,j{+}1),\ (i,j,k,k{+}1)\}\]
with $i{+}1<j$ {\it etc.}. Explicit computation shows that this $A_3$ gives an exactly same alphabet as 
\begin{equation}\label{6ptalphabet}
\{u,v,w,1{-}u,1{-}v,1{-}w,y_u,y_v,y_w\},\ y_u=\frac{1{+}u{-}v{-}w{+}\sqrt{(1{-}u{-}v{-}w)^2{-}4uvw}}{1{+}u{-}v{-}w{-}\sqrt{(1{-}u{-}v{-}w)^2{-}4uvw}}
\end{equation}
with 
\[u=\frac{\langle\bar ij\rangle\langle i\bar j\rangle\langle k{-}1\ k\ i{-}1\ i\rangle\langle k{-}1\ k\ j{-}1\ j\rangle}{\langle i(i{-}1\ i{+}1) (j{-}1\ j)(k{-}1\  k)\rangle\langle j(j{-}1\ j{+}1) (i{-}1\ i)(k{-}1\ k)\rangle}\]
and its cyclic images $v=u(i\to j,j\to k,k\to i)$, $w=u(i\to k,k\to j,j\to i)$. Especially, taking $i\to6$, $j\to 2$, $k\to 4$, this expression and its cyclic images go back to $u=\frac{\langle6123\rangle\langle3456\rangle}{\langle6134\rangle\langle2356\rangle}$, which is exactly the original cross-ratios definition for $n=6$. So we conclude that
\begin{tcolorbox}
The alphabet for two-loop $n$-point MHV amplitudes, \eqref{MHVsingularities}, always has cluster-algebraic structure as the union of $A_3$ as in \eqref{6ptalphabet}.
\end{tcolorbox}

\begin{figure}[htbp]
    \centering
    \includegraphics[width=0.35\linewidth]{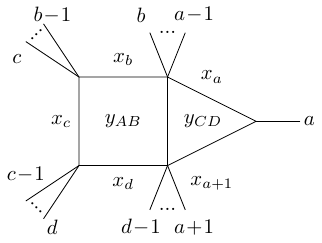}
    \caption{A box-triangle subsector with general external legs appearing in the $\mathrm{NMHV}$ Landau diagram.}
    \label{fig:nmhvsub}
\end{figure}

\paragraph{A preview for two-loop NMHV} At last, we note here that though we have only discussed MHV alphabet in above examples, similar analysis can be directly applied to $\mathrm{NMHV}$ and $\mathrm{N}^{2}\mathrm{MHV}$. The corresponding Landau diagrams have been determined in \cite{Prlina:2017tvx} and we can apply the Landau-based Schubert analysis to them. Let us take a simple example in $\mathrm{NMHV}$, which is depicted in Fig.~\ref{fig:nmhvsub} and show that the odd letters will appear just in the same way as we discussed before.
The Landau equations for $y_{AB}$ will be
\begin{equation}
    {\color{lightred}\mathrm{L}_{abcda+1}:}\,\,\langle ABa a+1\rangle=\langle AB a-1a\rangle=\langle AB b-1b\rangle=\langle AB c-1c\rangle=\langle AB d-1d\rangle=0.
\end{equation}
It can be easily seen that the following two Schubert configurations originating from two box-bubble subsectors of Fig.~\ref{fig:nmhvsub} can be combined to give ${\color{lightred}\mathrm{L}_{abcda+1}:}$
\begin{equation}
    \begin{aligned}
        {\color{lightblue}\mathrm{L}_{abcd}:}&\,\,\langle AB a-1a\rangle=\langle AB b-1b\rangle=\langle AB c-1c\rangle=\langle AB d-1d\rangle=0, \\
        {\color{lightyellow}\mathrm{L}_{a+1bcd}:}&\,\,\langle AB aa+1\rangle=\langle AB b-1b\rangle=\langle AB c-1c\rangle=\langle AB d-1d\rangle=0.
    \end{aligned}
\end{equation}
Two pairs of solutions in the one-dimensional subspace determined by $\langle AB b-1b\rangle=\langle AB c-1c\rangle=\langle AB d-1d\rangle=0$ are 
\begin{equation}
    \begin{aligned}
        l_{1,2}^{\pm}=&({c\!-\!1}c{b\!-\!1})\cap({d\!-\!1}d{b\!-\!1})+\alpha_{1,2}^{\pm}[({c\!-\!1}c{b\!-\!1})\cap({d\!-\!1}db)+({c\!-\!1}cb)\cap({d\!-\!1}d{b\!-\!1})] \\
        &+({\alpha_{1,2}^{\pm}})^{2}({c\!-\!1}cb)\cap({d\!-\!1}db),
    \end{aligned}
\end{equation}
where
\begin{equation}
    \begin{aligned}
        \alpha_{1}^{\pm}=&\frac{\langle{(c\!-\!1}c{b\!-\!1})\cap({d\!-\!1}db){a\!-\!1a} \rangle\langle{(c\!-\!1}cb)\cap({d\!-\!1}d{b\!-\!1}){a\!-\!1a} \rangle\pm \sqrt{\mathbf{Gram}_{a,b,c,d}}}{2\langle{(c\!-\!1}cb)\cap({d\!-\!1}db){a\!-\!1a} \rangle}, \\
        \alpha_{2}^{\pm}=&\frac{\langle{(c\!-\!1}c{b\!-\!1})\cap({d\!-\!1}db){aa\!+\!1} \rangle\langle{(c\!-\!1}cb)\cap({d\!-\!1}d{b\!-\!1}){aa\!+\!1} \rangle\pm \sqrt{\mathbf{Gram}_{a\!+\!1,b,c,d}}}{2\langle{(c\!-\!1}cb)\cap({d\!-\!1}db){aa\!+\!1} \rangle}.
    \end{aligned}
\end{equation}
Again, the distance between two solutions is
\begin{equation}
    (l_1\cdot l_2)=-\langle{d\!-\!1}d{b\!-\!1}b\rangle\langle{d\!-\!1}d{c\!-\!1}c\rangle\langle{b\!-\!1}b{c\!-\!1}c\rangle(\alpha_1-\alpha_2)^2.
\end{equation}
So the cross ratio can be calculated as
\begin{equation}
\begin{aligned}
    &\frac{(l_1^{+}\cdot l_{2}^{+})(l_1^{-}\cdot l_{2}^{-})}{(l_1^{+}\cdot l_{2}^{-})(l_1^{-}\cdot l_{2}^{+})}=\left[\frac{(\alpha_1^{+}-\alpha_{2}^{+})(\alpha_1^{-}-\alpha_{2}^{-})}{(\alpha_1^{+}-\alpha_{2}^{-})(\alpha_1^{-}-\alpha_{2}^{+})}\right]^2, \\
    &\frac{(\alpha_1^{+}-\alpha_{2}^{+})(\alpha_1^{-}-\alpha_{2}^{-})}{(\alpha_1^{+}-\alpha_{2}^{-})(\alpha_1^{-}-\alpha_{2}^{+})}=\frac{\mathbf{Gram}_{a,b,c,d}^{a+1,b,c,d}+\sqrt{\mathbf{Gram}_{a,b,c,d}\mathbf{Gram}_{a+1,b,c,d}}}{\mathbf{Gram}_{a,b,c,d}^{a+1,b,c,d}-\sqrt{\mathbf{Gram}_{a,b,c,d}\mathbf{Gram}_{a+1,b,c,d}}}.
\end{aligned}
\end{equation}
The gram determinants are defined as $\mathbf{Gram}_{\mathbf{a}}^{\mathbf{b}}=\det((x_{a_{i}}-x_{b_{j}})^2)$ and $\mathbf{Gram}_{\mathbf{a}}=\mathbf{Gram}_{\mathbf{a}}^{\mathbf{a}}$. Actually, once we have determined which Gram determinant appearing under the square root, above result can be generalized in a straightforward way. A thorough discussions for NMHV and N${}^2$MHV cases with new predictions for their alphabets will be presented in a forthcoming work~\cite{toappear}.

\subsection{Applications to two-loop form factors}
\begin{figure}[h]
    \centering
    \subfigure[planar topology]{\includegraphics[width=0.3\linewidth]{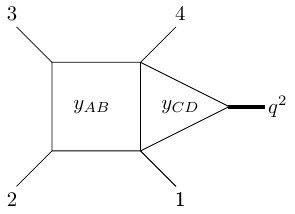}}
    \quad\quad\quad
    \subfigure[non-planar topology]{\includegraphics[width=0.25\linewidth]{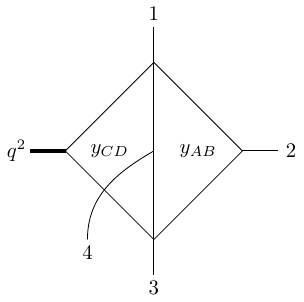}}
    \caption{Topologies considered for four-point form factor. We consider cut condition $\langle AB I_{\infty} \rangle=0$ and $\langle CD I_{\infty} \rangle=0$ as well. And we parameterize the massive leg $q^2$ with two massless legs $q_1$ and $q_4$, so two of the cut conditions for $(CD)$ represented by (a) are $\langle CD 4 q_4 \rangle=0$ and $\langle CD q_1 1 \rangle=0$.}
    \label{fig:FFtopo}
\end{figure}

In this subsection, we make an attempt to apply the method to form factors in ${\cal N}=4$ SYM which do not enjoy dual conformal invariance. Our Landau-based Schubert analysis would necessarily involve the so-called ``infinity twistor" $I_{\infty}$, which breaks conformal symmetry, as described in \cite{He:2022tph}. In this first attempt, let us present a state-of-art example of four-point MHV form factor at two loops, {\it c.f.}~\cite{Dixon:2022xqh,Guo:2024bsd} (the Schubert analysis for the three-point example was analyzed in terms of one-mass four-point Feynman integrals in~\cite{He:2022ctv}). Instead of studying all one-mass five-point Feynman integrals which would involve a huge number of Landau diagrams and Schubert configurations, we find simplifications similar to the amplitudes; it turns out that exactly two kinds of diagrams as shown in Fig.~\ref{fig:FFtopo} (plus their cyclic images) suffice to give all the Landau singularities and symbol letters after performing the corresponding Schubert analysis. Note that we do not have the guidance of the analog of ``amplituhedron" for form factors, thus we do not know how to choose Landau diagrams systematically. However, at least for four-point case, these two topologies are natural choices, since they are symmetric and have exactly the right number of propagators (after taking $\langle AB I_{\infty} \rangle=0$ and $\langle CD I_{\infty} \rangle=0$ into account), whose maximal cuts exactly localize the two loop momenta. 
As we will see later in this subsection, although some of the symbol letters constructed from Landau/Schubert analysis are not those for form factor, all of them do appear as letters of the 5-point one-mass Feynman integrals under consideration.

Before delving in the concrete analysis, let us briefly review the symbol letters of four-point form factor. For later convenience, we rewrite the symbol alphabet in \cite{Dixon:2022xqh} using usual Mandelstam variables as:
\begin{equation}
\begin{aligned}
   & \left\{ L_1 = q^2,\, L_2=s_{12},\ldots,\, L_6=s_{123},\ldots,\, L_{10}=q^2 -s_{123},\ldots,\right.\\
   & \quad \left. L_{14}=q^2(s_{12}-s_{123})+s_{123}(-s_{12}+s_{34}+s_{123}),\ldots\right\}_{\text{rational}}\cup\left\{L_{18},\ldots,L_{35}\right\}_{\text{algebraic}}
\end{aligned}
\end{equation}
The ellipses denote the image under cyclic of external legs $\{1,2,3,4\}$.

Now let us firstly consider the planar topology. Note that although the topology of the diagram shows up as a box-triangle apparently, since cut conditions $\langle AB I_{\infty} \rangle=0$ and $\langle CD I_{\infty} \rangle=0$ are considered as well, the top sector actually reads a penta-box integral, which is similar to the discussion for 2-loop MHV amplitudes. Analogous to that discussion, after analyzing all kinds of its subtopologies, we have Landau loci in the form of 4- or 5-point Gram determinants formed by dual points $\{(q_11),(12),(23),(34),(4q_4),I_\infty\}$, which (together with their cyclic image) take care of the rational letters $L_2 \sim L_{17}$.

Similar to previous discussion about amplitudes, we consider the constraints given by a specific (sub-)diagram, and ask whether they can be taken as combinations of other diagrams. If a diagram gives 6 constrains, it can be thought as combination of 3 Schubert configurations. And if a diagram gives 5 constrains, it can be thought as combination of 2 Schubert configurations. Projecting the momentum solution to momentum twistor of external lines, we will have many $A_n$'s on each of them.

Let us firstly consider a subsector, where we don't take $\langle CD I_\infty \rangle$ into account, the constraints obtained by Landau equation on $(AB)$ will be
\begin{equation}\label{eq:FFplanarCondition2}
    \begin{aligned}
        & \text{cut:} \quad \langle AB12 \rangle = \langle AB23 \rangle = \langle AB34 \rangle = \langle ABI_\infty \rangle =0,\\
        & \text{pinch:} \quad \langle AB q_1 1 \rangle = \langle AB 4 q_4 \rangle = 0 \left( = \langle 4 q_4 q_1 1 \rangle \propto L_1 \right) .
    \end{aligned}
\end{equation}
This is the only configuration that gives 6 constraints on $(AB)$, which can be seen as combination of 3 different Schubert configurations. We found that only box configurations can be combined (\textit{i.e.} no constraint like the second line in Eq.~\eqref{eq:MHVtopConstraints} involved), and there are totally 16 different ways to combine those box configurations. After projecting to momentum twistor of external lines and collecting all the $A_n$'s, we are able to construct 29 independent letters in total:  $\{ W_1 \sim W_{9},W_{12} \sim W_{15}, W_{18}, W_{19}, W_{22} \sim W_{24}, W_{33}, W_{34}, W_{37}, W_{38}, W_{40}, W_{43}\sim W_{48}\}$, where we have adopted the definition of $W_i$ in \cite{Abreu:2020jxa}.

Analysis for the top sector gives the following constraints on $(AB)$:
\begin{equation}\label{eq:FFplanarCondition}
    \begin{aligned}
        & \text{cut:} \quad \langle AB12 \rangle = \langle AB23 \rangle = \langle AB34 \rangle = \langle ABI_\infty \rangle =0,\\
        & \text{pinch:} \quad \langle ABq_1 1 \rangle \langle 4 q_4 I_\infty \rangle - \langle AB 4 q_4 \rangle \langle q_1 1 I_\infty \rangle = 0 
    \end{aligned}
\end{equation}
These 5 constraints can be thought of combination of the following two Schubert configuration
\begin{equation}
    \begin{aligned}
        \langle AB q_1 1 \rangle = \langle AB 12 \rangle = \langle AB 34 \rangle = \langle ABq_1 1 \rangle \langle 4 q_4 I_\infty \rangle - \langle AB 4 q_4 \rangle \langle q_1 1 I_\infty \rangle = 0,\\
       \langle AB I_\infty \rangle = \langle AB 12 \rangle = \langle AB 34 \rangle = \langle ABq_1 1 \rangle \langle 4 q_4 I_\infty \rangle - \langle AB 4 q_4 \rangle \langle q_1 1 I_\infty \rangle = 0.\\
    \end{aligned}
\end{equation}
Project on $(12)$ or $(34)$, we will have an $A_1$, and the letters read $\left\{ W_{36}\sqrt{\frac{W_{23} W_{24}}{W_{2}W_{7}}}, W_{23} W_{24} \right\}$. There are other 10 diagrams that give 5 constraints on $(AB)$, but all of them don't contain non-trivial constraints like the second line in Eq.~\eqref{eq:FFplanarCondition}. So we don't get new symbol letters from these configurations.

Notice the following relation:
\begin{equation}
\begin{aligned}
    L_{18}=W_{47},L_{20}=W_{34},L_{21}=W_{33}, L_{24}=1/\left(W_{37}W_{38}\right),\\
    L_{29}=W_{43}/W_{45},
    L_{30}=W_{46}/W_{44},L_{34}=W_{40},L_{35}=W_{43}W_{57}/W_{44},\\
    \left\{ W_{18},W_{20},W_{21},W_{24} \right\} \stackrel{\text{cyclic}}{\longrightarrow} \left\{ 1/W_{19},W_{23},W_{22},W_{25} \right\}.
\end{aligned}
\end{equation}
Then we have $\left\{ L_1\sim L_{25},L_{29},L_{30},L_{34},L_{35} \right\}$ at hand by analyzing the planar diagram.


Then we move on to the non-planar diagram. Similar to the discussion in \cite{He:2022tph}, we firstly solve the $(AB)$ cut condition,
\begin{equation}
    \begin{aligned}
        & \langle AB12 \rangle = \langle AB 23 \rangle =  \langle AB CD \rangle = 0\\
        &\langle AB4q_4 \rangle \langle CD I_\infty \rangle \langle 34 I_\infty \rangle - \langle AB34 \rangle \langle CD I_\infty \rangle \langle 4 q_4 I_\infty \rangle \\
        & - \langle ABI_\infty \rangle \langle CD 4q_4 \rangle \langle 34 I_\infty \rangle + \langle ABI_\infty \rangle \langle CD 34 \rangle \langle 4q_4 I_\infty \rangle = 0,
    \end{aligned}
\end{equation}
which will lead to a Jacobian concerning $(CD)$ (this Jacobian can also be seen as the pinch condition from $(AB)$ ), having several factors; then we cut two of the Jacobian factors and two original propagator $\langle CDq_1 1 \rangle$ and $\langle CD 4q_4  \rangle$, we'll get 2 intersection on $(4q_4)$. Mixed with box configurations, cross-ratios on $(4q_4)$ (together with their cyclic image) will give the rest of the letters $\left\{ L_{26} \sim L_{28}, L_{31} \sim L_{33} \right\}$.



\section{Conclusion and Discussions}

In this paper, we have made progress on the study of singularities and symbol letters for Feynman integrals by revisiting the Schubert analysis from the point of view of Landau analysis. First of all, we have put this conjectural method on a firm ground by showing that symbol letters computed as cross-ratios of intersecting points on lines of cut diagrams are consistent with Landau equations. On one hand, Landau analysis selects preferred geometries in twistor space (or embedding space in more general cases) on which we perform Schubert analysis. On the other hand, our method can be viewed as ``uplifting” Landau singularities to symbol letters; we have shown how Schubert analysis allows one to compute especially those algebraic letters which contain more information than the Landau singularities (that are polynomials in twistor variables). 
We have also discussed how this can be extended to Feynman integrals that evaluate to elliptic MPL functions {\it etc.}. 

We have then applied the method to scattering amplitudes and form factors in ${\cal N}=4$ SYM theory, where a very small number of Landau diagrams suffice to determine the full alphabet of {\it e.g.} two-loop $n$-point MHV amplitudes and form factors. Among other things, our method explains why the symbol alphabet often turns out to be (union of) cluster algebras since Schubert analysis always gives a collection of inter-related type-$A$ cluster algebras; {\it e.g.} for $n=6,7$ they become $A_3$ and $E_6$ cluster algebras respectively. 
We believe that the connections between Landau analysis and Schubert analysis revealed in this paper will have implications for both approaches and related studies. Let us briefly comment on directions for future investigations. 

First of all, we would like to ask what are the most general settings for applying Landau-based Schubert analysis. By using the Schubert analysis in embedding formalism, one can apply this method to integrals in general dimensions and those with internal masses {\it etc.} and more importantly we should be able to generalize this method to the Baikov representation for most general Feynman integrals~\cite{He:2023umf,Jiang:2024eaj}. In addition, although we have mostly restricted ourselves to MPL functions, it would be extremely interesting to apply it for Feynman integrals that evaluate to elliptic MPL or even more complicated cases~\cite{Bourjaily:2019hmc}. The Landau analysis of course still determines possible singularities, but the corresponding Schubert analysis already becomes more subtle for elliptic cases (and even more so beyond), and it would be fascinating to understand the meaning of these ``letters" which contain more information than the locations of singularities. For all these purposes, it would be important to develop the automated package more systematically. 

On the other hand, we have only scratched the surfaces of computing the complete alphabet of {\it physical quantities} using this method, even just for the special case of ${\cal N}=4$ SYM theory. It is straightforward to apply this method to two-loop amplitudes with higher $k$, {\it e.g.} we have obtained predictions for the alphabets of $n$-point NMHV amplitudes (and even some higher $k$ results) based on Landau diagrams given in~\cite{Prlina:2017azl,Prlina:2017tvx}, and we expect this to work exactly the same way for higher-loop amplitudes (as long as one can determine the relevant Landau diagrams). Our method also successfully produce the alphabets for four-point form factors using only two Landau diagrams, and it would be highly desirable to generalize this analysis to higher points or higher loops. To proceed, we would need to determine the relevant Landau diagrams which contain all possible singularities of the form factor, and then make predictions for the alphabets. It is also tempting to wonder if one could ``select" certain Landau diagrams (as opposed to compute the alphabets of all relevant Feynman integrals) which could account for the complete alphabet of amplitudes {\it etc.} in more general theories like QCD.

Last but not least, our method has provided a nice explanation for the observation that symbol alphabets for physical quantities and those for individual Feynman diagrams seem to be closely related to cluster algebras. From a mathematical point of view, already for ${\cal N}=4$ SYM this amounts to a map from any Landau diagram to a union of type-$A$ cluster algebras embedded in Gr$(4,n)$, which {\it e.g.} for two-loop MHV case combine to $A_3$ and $E_6$ for $n=6,7$ but goes beyond any finite-type cluster algebra for $n\geq 8$.  It would certainly be very interesting to explore this connection further.

\section*{Acknowledgement}
The work of SH is supported by the National Natural Science Foundation of China under Grant No. 12225510, 11935013, 12047503, 12247103, and by the New Cornerstone Science Foundation through the XPLORER PRIZE. Q.Y. is funded by the European Union (ERC, UNIVERSE PLUS, 101118787). Views and opinions expressed are however those of the author(s) only and do not necessarily reflect those of the European Union or the European Research Council Executive Agency. Neither the European Union nor the granting authority can be held responsible for them.

\appendix

\section{The leading Landau singularities and pinch conditions from loop-by-loop analysis}\label{app:triangle}
In the main text, we usually solved the Landau equations for higher-loop integrals from a loop-by-loop viewpoint. The general idea is as the following: we choose a special order of all loop momenta and solve their Landau equations one by one. On the support of cut conditions, pinch condition of a loop momentum always results in new cut conditions for the rest loop momenta, reducing the Landau equation system to lower loops recursively. Leading Landau locus of the integral then reads the locus from Landau equation for the last loop momentum. For instance, when looking into the box-bubble integral Fig.~\ref{fig:6pdbsub_bb}, we firstly concentrate on $y_{CD}$, whose Landau equations read (we emphasize again we do not consider second-type Landau loci in this note)
\begin{align}
  &\alpha_1\langle CD61\rangle=\alpha_2\langle CDAB\rangle=0\nonumber\\
  &\alpha_1(y_{CD}-x_{1})^\mu+\alpha_2(y_{CD}-y_{AB})^\mu=0
\end{align}
On the support of cut conditions, the pinch condition is equivalent to 
\begin{equation}
    \left(\begin{matrix}
        0&\langle AB 61\rangle\\ \langle AB61\rangle &0    \end{matrix}\right)\cdot\left(\begin{matrix}
        \alpha_1\\ \alpha_2
    \end{matrix}\right)=0
\end{equation}
{\it i.e.} $\langle AB61\rangle=0$. Therefore now for $y_{AB}$ we have a box configuration equivalently, and we conclude that generally for every bubble sub-diagram in an $L$-loop Feynman integral, it always results in an $(L{-}1)$-loop Schubert configuration with bubble sub-diagram replaced by a propagator. 

\begin{figure}
    \centering
    \includegraphics[width=0.3\linewidth]{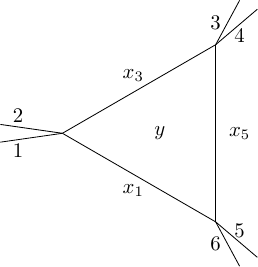}
    \caption{One-loop triangle with all external legs massive}
    \label{fig:triangle}
\end{figure}

Similar discussion also applies to  leading singularities for triangles. However, here is one point we should emphasize. The pinch condition for a triangle in Fig.~\ref{fig:triangle} will be
\begin{equation}
    \left(\begin{array}{ccc}
        0 & \langle6123\rangle & \langle6145\rangle \\
        \langle6123\rangle & 0 & \langle2345\rangle \\
        \langle6145\rangle & \langle2345\rangle & 0
    \end{array}\right)\cdot
    \left(\begin{array}{c}
         \alpha_{1} \\ \alpha_{3} \\ \alpha_{5}
    \end{array}\right)=0
\end{equation}
For existence of solutions, we would conclude that $\langle6123\rangle\langle6145\rangle\langle2345\rangle=0$ following from a similar logic. However, if we require to solve the leading singularity of the triangle, {\it i.e.} every $\alpha_{i}\neq 0$, then we should instead set
\begin{equation}    \langle6123\rangle=\langle6145\rangle=\langle2345\rangle=0
\end{equation}
This constraint is stronger than $\langle6123\rangle\langle6145\rangle\langle2345\rangle=0$. Similarly, once there is a triangle in higher-loop diagram, the leading-singularity condition (every $\alpha_{i}\neq 0$) will result in three conditions for the corresponding loop. This is exactly what we did in solving Landau equation of {\it e.g.} Fig.~\ref{fig:6pdbsub_bt}.

Note that similar things happen for more complicated topologies by requiring \textit{all} parameters not being 0. For example, for the top sector of the double-box we studied in Fig.~\ref{fig:doublebox}, the pinch condition for $y_{CD}$ will give us the following system
\begin{equation}
    \left(\begin{array}{cccc}
        0 & 0 & 0 & \langle AB61\rangle \\
        0 & 0 & \langle1256\rangle & \langle AB12\rangle \\
        0 & \langle1256\rangle & 0 & \langle AB56\rangle \\
        \langle AB61\rangle & \langle AB12\rangle & \langle AB56\rangle & 0
    \end{array}\right)\cdot
    \left(\begin{array}{c}
         \alpha_{1} \\ \alpha_2 \\ \alpha_6 \\ \beta
    \end{array}\right)=0.
\end{equation}
Requiring non-zero solution for all four parameters $\alpha_{1},\alpha_{2},\alpha_{6},\beta$, we will derive the following constraints
\begin{equation}
    \langle AB61\rangle=\langle AB12\rangle=\langle AB56\rangle=\langle 1256\rangle.
\end{equation}
instead of only determinant of the matrix reading 0. Combining this with the cut condition for $y_{AB}$ will give us:
\begin{equation}\label{eq:sixconstraints}
    \langle AB23\rangle=\langle AB34\rangle=\langle AB45\rangle=\langle AB61\rangle=\langle AB12\rangle=\langle AB56\rangle=0.
\end{equation}
This is the most constrained condition in this problem, which brings us the $A_3$ structure.

\bibliographystyle{utphys}
\bibliography{ref}

\begin{landscape}
\begin{figure}
    \centering
    \includegraphics[width=\linewidth]{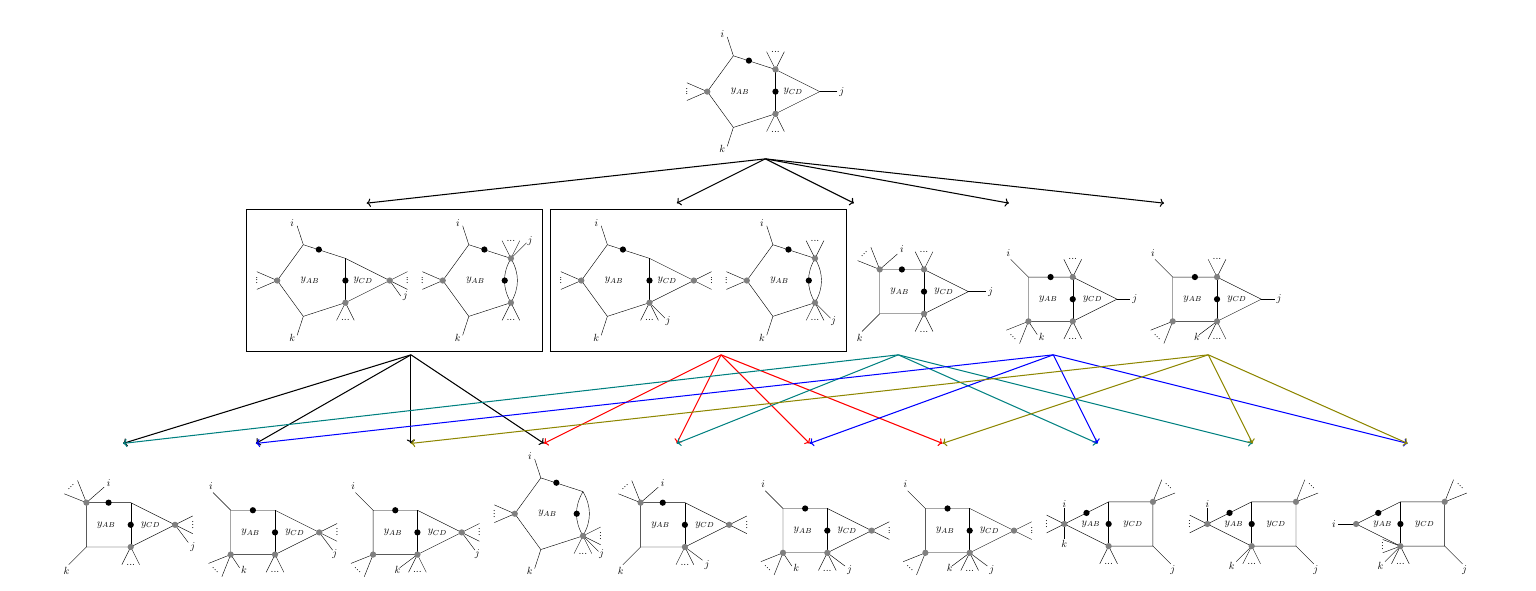}
    \caption{Landau diagrams for 2-loop MHV amplitudes II}
    \label{fig:MHV2}
\end{figure}
\end{landscape}

\end{document}